\documentclass[12pt]{article}
\usepackage{amsmath, amssymb}
\usepackage{amsthm}
\usepackage{graphicx}
\usepackage{enumerate}
\usepackage{url} 

\usepackage[utf8]{inputenc} 
\usepackage[english]{babel}
\usepackage{csquotes}

\numberwithin{equation}{section}
\counterwithin{table}{section}

\newcommand{\blind}{1}

\usepackage{booktabs} 
\usepackage{tabularray} 
\usepackage{tablefootnote}
\usepackage{float} 
\floatstyle{plaintop}
\restylefloat{table}

\IfFileExists{xurl.sty}{\usepackage{xurl}}{} 

\usepackage{graphicx} 
\usepackage[outdir=./]{epstopdf}

\usepackage{algorithm}
\usepackage{algpseudocode}

\usepackage{multirow} 
\usepackage{array} 
\usepackage{etoolbox} 
\usepackage{paralist} 
\usepackage{verbatim} 
\usepackage{subfig} 
\usepackage[colorlinks=true,citecolor=blue,linkcolor=blue]{hyperref} 

\usepackage{tikz} 
\usepackage{xcolor,color}
\usetikzlibrary{fit}

\renewcommand{\labelenumii}{\textbf{A\arabic{enumii}}}

\theoremstyle{remark}
\newtheorem{remark}{Remark}

\theoremstyle{plain}
\newtheorem{theorem}{Theorem}

\usepackage[
authordate,
giveninits=true,
uniquename=false,
backend=biber,
natbib=true,
doi=only,
isbn=false,
url=false,
maxcitenames=2,
noibid
]{biblatex-chicago}

\NewBibliographyString{available}
\NewBibliographyString{urlwhen}

\DefineBibliographyStrings{english}{%
  available = {Available at},
}

\DefineBibliographyStrings{english}{%
  urlwhen = {accessed},
}

\begin{document}

\def\spacingset#1{\renewcommand{\baselinestretch}%
{#1}\small\normalsize} \spacingset{1}

\if1\blind
{
  \title{\bf Overlap-Excluded Estimation of Participation Probabilities for Non-Probability Samples}
  \author{Lyndon Ang, Robert Clark, Bronwyn Loong and Anders Holmberg\footnote{Lyndon Ang, College of Business and Economics, Australian National University, Canberra, ACT, 2600, Australia and Australian Bureau of Statistics, Belconnen, ACT, 2617, Australia. Email: lyndon.ang@abs.gov.au; Robert Clark, College of Business and Economics, Australian National University, Canberra, ACT, 2600, Australia. Email: robert.clark@anu.edu.au; Bronwyn Loong, College of Business and Economics, Australian National University, Canberra, ACT, 2600, Australia. E-mail: bronwyn.loong@anu.edu.au; Anders Holmberg, Australian Bureau of Statistics, Belconnen, ACT, Australia. E-mail: anders.holmberg@abs.gov.au. The views expressed in this paper are those of the authors and do not necessarily represent the views of the Australian Bureau of Statistics. Where quoted or used, they should be attributed clearly to the authors.}}

  \date{}
  
  \maketitle
} \fi

\if0\blind
{
  \bigskip
  \bigskip
  \bigskip
  \begin{center}
    {\LARGE\bf Overlap-Excluded Estimation of Participation Probabilities for Non-Probability Samples}
\end{center}
  \medskip
} \fi

\begin{abstract}

    Correction of selection bias is crucial for making valid statistical inferences about a finite population using non-probability datasets.  Various approaches have been proposed to correct this bias and obtain estimates of finite population quantities such as means and totals.  The quasi-randomisation approaches, in particular, aim to estimate a `participation probability' for units in the non-probability dataset.  These probabilities are then applied in an inverse probability weighting scheme to produce population estimates.  In this paper, we propose to estimate the participation probabilities for the non-probability sample by combining the non-probability sample with an available probability sample, while excluding the overlapping units in the probability sample.  When the true overlap cannot be determined, we show how the Expectation-Maximization method can be used to simultaneously estimate the overlap and the participation probabilities.  We compare the performance of our approach against several alternative methods theoretically and in an empirical context, and demonstrate the efficiency gains that are possible when the overlapping probability sample is excluded.

\end{abstract}

\noindent%
\vfill


\newpage

\spacingset{1.8} 
\setlength{\parindent}{0pt}
\setlength{\parskip}{6pt plus 2pt minus 1pt}

\section{Introduction} \label{sec:intro}

Until recent times, the sample survey paradigm has revolved around the collection of variables of interest from a carefully designed probability sample of the population using a well-designed questionnaire or other instrument.  In this context, administrative datasets have been seen as useful to provide auxiliary variables which can modestly reduce the variances of estimators, for example through calibrated weighting in a model-assisted framework (\citealt{sarndal.etal_1992}).

As a byproduct of the ongoing digitalisation of society, there is now an increasing amount of data being collected about individuals, entities and the environment.  Some examples of this data include mobile device location information, satellite imagery, commercial transactions, administrative byproducts, and web panel surveys (see for example \citealt{tam.vanhalderen_2020}, \citealt{beaumont_2020}, and \citealt{tam.clarke_2015}).  Some of these datasets now contain direct measurements of variables of interest once collected only through a sample survey.  However, many of these datasets are primarily generated for non-statistical purposes, and so may not accurately represent the population of interest.  In particular, they may suffer from substantial missingness, and particularly selection bias where some subgroups of the target population are less likely to be included in the dataset than others.  When this occurs, they may be better considered non-probability samples.  Using these datasets directly without adjustment to produce inferences about population parameters such as totals and means is likely to result in biased estimates (see for example \citealt{bethlehem_2010} and \citealt{meng_2018}).

Various approaches have been developed to address selection bias in non-probability data.  \citet{zhang_2019}, \citet{rao_2021}, and \citet{wu_2022a} review developments in this area, while \citet{ang.etal_2025} undertook an empirical comparison of different methods for addressing selection bias in large non-probability datasets.  In these approaches, the probability samples play a new role to help address selection bias in the non-probability data.  In this new paradigm, the probability sample is very much the support act and not the lead, and may not collect the variables of interest at all.

Inference approaches vary depending on the data items available on the probability and non-probability datasets.  They may also be grouped into three broad categories: imputation-based approaches, where the variable of interest is imputed for the dataset it is not available in (see for example \citealt{rivers_2007} and \citealt{medous.etal_2023a}); weighting-based approaches which create a weight associated with each record on the probability or non-probability sample (see for example \citealt{kim.tam_2021} and \citealt{elliott.valliant_2017}); and doubly- or multiply-robust approaches which combine weighting and imputation (see for example \citealt{chen.etal_2020} and \citealt{haziza.chen_2022}).

In this paper the focus is on the weighting-based quasi-randomisation approach to addressing selection bias.  The approach assumes that the missingness in the non-probability sample is governed by an unknown selection mechanism that may be estimated via a model using available auxiliary information.  The estimated model is applied to the units in the non-probability data to produce estimated propensities of selection, also known as participation probabilities.  The inverse of these propensities are used as weights to form an Inverse Probability Weighted (IPW) estimator for the non-probability sample to estimate finite population quantities.

This paper focusses on the case when the variable of interest is available on the non-probability dataset only, and auxiliary data items are available on both the non-probability dataset and a separate probability sample, also known as a reference sample.  This scenario may apply, for example, in the case of a probability sample of farms collecting general information about farm characteristics, alongside incomplete-coverage administrative records relating to specific production activities of farms.  Another example is where a non-probability dataset on specific health topics may be supplemented with a probability sample relating to general health characteristics.

\citet{valliant.dever_2011}, \citet{elliot_2009} (see also \citealt{elliott.valliant_2017}) and \citet{wang.etal_2021} pooled the auxiliary data from the non-probability and probability samples to perform their model fitting.  \citet{chen.etal_2020} proposed a pseudo-likelihood approach which does not require the samples to be pooled together for the model fitting.  \citet{burakauskaite.ciginas_2023} looked at the \citet{chen.etal_2020} approach for the case when auxiliary data for the full population is available to assist with estimating the participation probabilities.  These methods do not require any linkage of the data from the two samples.

In some scenarios, it may be possible to identify the overlap between the reference sample and the non-probability dataset.  This would occur, for example, if a common unit identifier is available (for example, a unique business or person number), through the addition of questions on one sample asking about membership in the other, or through linking using common matching variables.  Some approaches take advantage of this information to estimate the propensities.  By linking the indicator of membership in the non-probability dataset to the probability sample, \citet{kim.wang_2019} maximised a pseudo-likelihood based on the probability sample alone to estimate the parameters of the participation probability model.  \citet{liu.etal_2022} extended the \citet{elliott.valliant_2017} method by using the units present in one of the two samples but not both.  \citet{beaumont.etal_2024b} proposed a sample likelihood approach which uses the pooled sample but excludes duplicate units from the overlapping sample to estimate the participation probability model parameters.   \citet{kim.kwon_2024} proposed a similar approach which they referred to as conditional likelihood, again using the pooled sample but excluding the duplicate units from the overlapping sample.

This paper contributes to the literature on inference for non-probability samples.  Building on the pioneering work of previous authors, we propose an Overlap-Excluded (OE) pseudo-likelihood approach to estimating participation probabilities when the overlap with a reference sample is known.  Our other major contributions include a theoretical and empirical comparison of the efficiency of population estimates produced using the OE approach relative to existing methods.  We show theoretically, under the assumption that the reference sample is a Poisson sample, that the OE approach is always more efficient compared to the methods of \citet{chen.etal_2020} and \citet{kim.wang_2019}.  We provide scenarios when the OE approach is expected to perform better or worse than the \citet{beaumont.etal_2024b} and \citet{wang.etal_2021} methods.

Further, a simulation using real business taxation data illustrates the behaviour of competing methods under skewed populations and different reference sample sampling schemes.  When it is not possible to identify the overlapping units, we show how to employ the Expectation-Maximisation (EM) algorithm to simultaneously estimate the overlap and produce the maximum pseudo-likelihood estimator of the participation probability model parameters.  Empirical comparisons of these two approaches with other methods under simulation settings with both synthetic and real data demonstrate that the OE approaches outperform other pseudo-likelihood approaches when the non-probability dataset is large, while the sample likelihood approach of \citet{beaumont.etal_2024b} generally performs slightly better than the OE approaches.  An exception to this is the case when the inclusion probabilities of units into the reference sample depends on their selection into the non-probability sample - for this case, the OE method had slightly better performance.

The remainder of the paper is structured as follows.  We outline the basic setup for the paper in Section \ref{setup}.  Section \ref{oe_propscores} introduces the Overlap-Excluded propensity score estimator which is applicable when the overlap between the probability and non-probability samples can be determined, and compares its efficiency with several alternative methods under the assumption that the reference sample is a Poisson sample.  Section \ref{em_approach} describes the use of the EM algorithm to simultaneously estimate the overlapping sample and participation probabilities when the overlap between the non-probability sample and reference sample is not known.  In Section \ref{simstudy} we compare the performance of our estimators through a simulation study.  We apply the methods in a simulation using real-world data in Section \ref{real_data}.  Finally, Section \ref{conclusion} provides some concluding remarks.

\section{Basic Setup} \label{setup}

Let $\mathcal{U}$ be a finite population of size $N$.  For each unit $i$ in the population, we have values $(\boldsymbol{x}_i,y_i)^T$ for a data item of interest $y$ and auxiliary data items $\boldsymbol{x}$.  In this paper we focus on producing estimates of the population mean for $y$, defined as $\mu_y = \bar{Y}=N^{-1}\sum_{i=1}^{N}y_i$.  However, the proposed approach also applies to other quantities such as population totals.

Suppose we have a probability sample $A$ (which we will call the ``reference'' sample) of size $n_A$ drawn from the population.  $A$ contains information for $\boldsymbol{x}$ but not $y$.  Let $I_i$ be an indicator variable for unit $i$ being included in the sample $A$.  Define $\pi^A_i=P(i \in A) = P(I_i=1|\boldsymbol{q}_i)$ as the inclusion probability for unit $i$ to be selected in the probability sample, where $\boldsymbol{q}$ are the design variables governing inclusion into $A$.  Further, define $d^A_i=1/\pi^A_i$ as the survey design weight for $i \in A$.  The $\pi^A_i$ and $d^A_i$ values are known from the sample design.  In order to simplify working, in this paper we assume there is full response in $A$.  In practice, though, $A$ will generally have some non-response.  To address potential non-response bias, the initial weights may be adjusted through an explicit non-response adjustment which accounts for the propensity to respond (see for example Section 9.2 of \citealt{wu.thompson_2020}).  Alternatively, calibrated survey weights (\citealt{deville.sarndal_1992}) may be calculated using auxiliary population benchmarks to account for non-response bias and to improve efficiency.  Subject to assumptions (particularly that non-response is ignorable conditional on the benchmark variables), the inverse of the calibrated weights are estimates of the combined selection and response probability and can be used in place of $\pi^A_i$; see, for example, Chapter 6 of \citet{sarndal.lundstrom_2006}.

We also have a non-probability sample $B$.  $B$ contains information for $\boldsymbol{x}$ as well as $y$.  Let $\delta_i=I(i \in B)$ be an indicator variable for unit $i$ being included in the sample $B$.  The non-probability sample size is $N_B=\sum_{i=1}^{N}{\delta_i}$.  In contrast to $\pi^A_i$ in the reference probability sample, $\pi^B_i=P(\delta_i=1|\mathcal{U})$ is unknown and needs to be estimated.

Let $C=\mathcal{U} \setminus B$ be the units in the population not included in $B$, while the overlapping units are given by $D = A \cap B$.  There may or may not be overlap between the two samples $A$ and $B$.

We make some assumptions regarding the selection mechanism for units being included in $B$, in order to facilitate forming inferences using those datasets; see for example \citet{chen.etal_2020}.  These assumptions are:
\begin{enumerate}
	\item[]
    \begin{enumerate}
        \item \textit{Ignorability}: Conditional on the set of covariates $\boldsymbol{x}_i$, $\delta_i$ and $y_i$ are independent.
        \item \textit{Positivity}:  Conditional on $\boldsymbol{x}_i$, $P(\delta_i=1|\boldsymbol{x}_i) > 0, \forall i \in \mathcal{U}$.
        \item \textit{Independence}: Conditional on $\boldsymbol{x}_i$ and $\boldsymbol{x}_j$, $\delta_i \perp \delta_j, i \neq j, \forall i,j \in \mathcal{U}$.
    \end{enumerate}
\end{enumerate}

Ignorability in \textbf{A1} implies that $P(\delta_i=1|\boldsymbol{x},y)=P(\delta_i=1|\boldsymbol{x})$.  In other words, selection into $B$ is ignorable conditional on the covariates $\boldsymbol{x}$.  This assumption is similar to the Missing-At-Random (MAR) scenario of \citet{rubin_1976}.  \citet{andridge.etal_2019} and \citet{boonstra.etal_2021} refer to this type of selection process as Selection At Random (SAR).  In this paper, we adopt the SAR terminology to distinguish against the missing-data/nonresponse scenarios in survey sampling, and to reflect that in most non-probability data cases the process for generating the final non-probability dataset can most appropriately be described as a \textit{selection} into the dataset.

The positivity assumption \textbf{A2} states that conditional on $\boldsymbol{x}$ every unit in the population has a non-zero chance of inclusion in $B$.  This may not always hold - for example, in a dataset containing data from a social media platform, only those persons who are members of the platform have a chance of inclusion in the dataset.

In addition to the above, we assume that it is possible to accurately identify which units in $A$ are also present in $B$.  This assumption may be satisfied if there exists a common unit identifier available on $B$ and the reference sample $A$, or if it is possible to accurately link records in $A$ to $B$ using a common set of linking variables.  Alternatively, this assumption could be met by including a question in the survey instrument for $A$ which asks whether the respondent is also on the dataset $B$.

\section{Overlap-Excluded Propensity Score Estimation}\label{oe_propscores}

Our end goal is to produce an estimate of the population mean using $B$.  This can be achieved by using an Inverse Probability Weighted (IPW) estimator such as that suggested by \citet{chen.etal_2020}:
\begin{equation} \label{eqn:IPW}
    \hat{\mu}_{IPW} = \frac{1}{\hat{N}_B}\sum_{i \in B}\frac{y_i}{\hat{\pi}^B_i}
\end{equation}
where $\hat{N}_B = \sum_{i \in B}(\hat{\pi}^B_i)^{-1}$.  The problem we address in this paper is how best to produce $\hat{\pi}^B_i$.

We assume that the $\pi^B_i$ are governed by a parametric model, so that $\pi^B_i=\pi(\boldsymbol{x}_i,\boldsymbol{\theta}_0)$, where $\boldsymbol{\theta}_0$ are the unknown (true) model parameters.  Under assumption \textbf{A3} the population likelihood function of $\boldsymbol{\theta}$ over $\mathcal{U}$ is
\begin{equation} 
    L(\boldsymbol{\theta}) = \prod_{i \in \mathcal{U}} (\pi(\boldsymbol{x}_i, \boldsymbol{\theta}))^{\delta_i}(1-\pi(\boldsymbol{x}_i, \boldsymbol{\theta}))^{(1-\delta_i)} \nonumber
\end{equation}

and the corresponding log-likelihood function is
\begin{equation} \label{eqn:loglikelihood}   
    l(\boldsymbol{\theta}) = \sum_{i \in \mathcal{U}}\delta_i \text{log}(\pi(\boldsymbol{x}_i, \boldsymbol{\theta})) + \sum_{i \in \mathcal{U}}(1 - \delta_i)\text{log}(1 - \pi(\boldsymbol{x}_i, \boldsymbol{\theta}))
\end{equation}

If $\boldsymbol{x}$ is available for the full population, for example from a census of the population or from a population frame, then $\hat{\boldsymbol{\theta}}$ can be produced directly by maximising $l(\boldsymbol{\theta})$ to yield the maximum likelihood estimator $\hat{\boldsymbol{\theta}}$.  The resulting model parameter estimates can then be used to produce estimated participation probabilities $\hat{\pi}^B_i = \pi(\boldsymbol{x}_i,\hat{\boldsymbol{\theta}})$.

The focus in this paper is on the case when $\boldsymbol{x}$ is not available for the full population.  The second term in (\ref{eqn:loglikelihood}) is then not observed.  However, we can expand the term, which yields an alternate expression for the log-likelihood:
\begin{equation} \label{eqn:loglikelihood_alt}
    l(\boldsymbol{\theta}) = \sum_{i \in \mathcal{U}}\delta_i \text{log}(\pi(\boldsymbol{x}_i, \boldsymbol{\theta})) + \sum_{i \in \mathcal{U}}\text{log}(1 - \pi(\boldsymbol{x}_i, \boldsymbol{\theta})) - \sum_{i \in \mathcal{U}}\delta_i\text{log}(1 - \pi(\boldsymbol{x}_i, \boldsymbol{\theta})).
\end{equation}

It can be seen from (\ref{eqn:loglikelihood_alt}) that the first and third terms of the log-likelihood can be obtained directly from $B$.  \citet{chen.etal_2020} estimate the second term using the reference sample $A$.  Their approach, which we label as CLW, has a resulting pseudo-likelihood given by
\begin{equation} \label{eqn:chenetal.pl}
    l_{CLW}(\boldsymbol{\theta}) = \sum_{i \in B}\text{log}\frac{\pi(\boldsymbol{x}_i, \boldsymbol{\theta})}{1 - \pi(\boldsymbol{x}_i, \boldsymbol{\theta})} + \sum_{i \in A}d^A_i \text{log}(1 - \pi(\boldsymbol{x}_i, \boldsymbol{\theta})).
\end{equation}

Note that the second term in (\ref{eqn:loglikelihood}) is actually a sum over the population units that are not in $B$.  Under the assumption that it is possible to accurately identify which units in $A$ are also in $B$, the second term in (\ref{eqn:loglikelihood}) may alternatively be directly estimated using $A \setminus B$.  The resulting pseudo likelihood is given by
\begin{equation} \label{eqn:new.loglikelihood}   
    l_{OE}(\boldsymbol{\theta}) = \sum_{i \in B}\text{log}\pi(\boldsymbol{x}_i, \boldsymbol{\theta}) + \sum_{i \in A \setminus B}d^A_i \text{log}(1 - \pi(\boldsymbol{x}_i, \boldsymbol{\theta})).
\end{equation}

The design expectation of (\ref{eqn:new.loglikelihood}) conditional on $B$ will be equal to (\ref{eqn:loglikelihood}) if $\pi^A_i = P(i \in A|B)$, or if the two samples $A$ and $B$ are conditionally independent, that is, $P(i \in A|B) = P(i \in A|\boldsymbol{q})$.  Conditional independence means that knowledge of $B$ contains no further information about the inclusion probabilities for $A$ once the design variables $\boldsymbol{q}$ are taken into account.

In this paper, we will refer to $l_{OE}(\boldsymbol{\theta})$ as the Overlap-Excluded (OE) pseudo-likelihood, as only the non-overlapping units from the reference sample are included in the second term of the pseudo-likelihood.  In a similar manner, we term the estimated participation probabilities $\pi(\boldsymbol{x}_i,\hat{\boldsymbol{\theta}})$ the OE probabilities or propensities, and the IPW estimates created using the OE probabilities as OE IPW estimates.

It is worth pointing out the connection between the CLW and OE approaches.  \citet{chen.etal_2020} estimate the quantity $\sum_{i \in \mathcal{U}}\text{log}(1 - \pi(\boldsymbol{x}_i, \boldsymbol{\theta}))$ in (\ref{eqn:loglikelihood_alt}) by the Horvitz-Thompson estimator $\sum_{i \in A}d^A_i\text{log}(1 - \pi(\boldsymbol{x}_i, \boldsymbol{\theta}))$.  However, it can be seen from (\ref{eqn:new.loglikelihood}) that the OE approach is essentially estimating that same quantity using
\begin{equation}
    \sum_{i \in B}\text{log}(1 - \pi(\boldsymbol{x}_i, \boldsymbol{\theta})) + \sum_{i \in A \setminus B}d^A_i\text{log}(1 - \pi(\boldsymbol{x}_i, \boldsymbol{\theta})). \nonumber
\end{equation}
This is the dual-frame estimator from \citet{kim.tam_2021}, which is expected to be more efficient than the Horvitz-Thompson estimator, especially when $B$ is large and when the overlap between $A$ and $B$ is significant.

\begin{remark}
    $l_{OE}(\boldsymbol{\theta})$ bears similarities to a number of other approaches proposed in the literature which also exclude the overlapping units.  \citet{kim.wang_2019} propose a data integration propensity score estimator, which we label as KW, with a pseudo-likelihood of the form
    \begin{equation} \label{eqn:KW.loglikelihood}   
        l_{KW}(\boldsymbol{\theta}) = \sum_{i \in A}d^A_i \delta_i \text{log}(\pi(\boldsymbol{x}_i;\boldsymbol{\theta}))+ \sum_{i \in A}d^A_i (1 - \delta_i)\text{log}(1 - \pi(\boldsymbol{x}_i;\boldsymbol{\theta}))
    \end{equation}
    where the first term of (\ref{eqn:KW.loglikelihood}) is estimated using $A$ instead of being computed directly from $B$, as we have proposed.  Their approach also requires knowledge of $\delta_i, i \in A$.  Similar to the OE approach, the KW method requires $A$ and $B$ to be conditionally independent so that knowledge of $B$ contains no further information about the inclusion probabilities for $A$ once design variables $\boldsymbol{q}$ are taken into account.  The sample likelihood (SL) approach described in \citet{beaumont.etal_2024b} utilises the sample $A \cup B$ containing the unique elements from each sample - duplicates from the overlapping sample are removed.  The approach makes use of the conditional probability $p_{si} = P(\delta_i | i \in A \cup B, \boldsymbol{x}_i) = \pi^B_i/(\pi^B_i + \pi^A_i - \pi^B_i\pi^A_i)$.  The log-likelihood of $\boldsymbol{\theta}$ for this approach is
    \begin{align}
        l_{SL}(\boldsymbol{\theta}) &= \sum_{i \in A \cup B} \Bigl[\delta_i \text{log}(p_{s}(\boldsymbol{x}_i; \boldsymbol{\theta})) + (1 - \delta_i)\text{log}(1 - p_{s}(\boldsymbol{x}_i; \boldsymbol{\theta})) \Bigr] \nonumber \\
        &= \sum_{i \in B} \text{log}(p_{s}(\boldsymbol{x}_i; \boldsymbol{\theta})) + \sum_{i \in A}\text{log}(1 - p_{s}(\boldsymbol{x}_i; \boldsymbol{\theta})) - \sum_{i \in A \cap B}\text{log}(1 - p_{s}(\boldsymbol{x}_i; \boldsymbol{\theta})) \label{eqn:SL.loglikelihood}
    \end{align}
    where $p_s(\boldsymbol{x}_i;\boldsymbol{\theta}) = \pi(\boldsymbol{x}_i;\boldsymbol{\theta})/(\pi(\boldsymbol{x}_i;\boldsymbol{\theta}) + \pi^A_i - \pi(\boldsymbol{x}_i;\boldsymbol{\theta})\pi^A_i)$.  The SL approach is unweighted and does not include a $d^A_i$ term.  The SL method assumes that $A \perp B$.
\end{remark}

\begin{remark}
    In contrast to the OE, CLW, KW and SL approaches, the Adjusted Logistic Propensity (ALP) approach described in \citet{wang.etal_2021} takes a ``stacked sample'' approach whereby all the sample units in $A$ and $B$, including overlapping sample, are stacked together.  Each element of the stacked sample is assumed to be distinct even though the overlapping sample will be present twice, once in $B$ and once in $A$.  The approach has the pseudo log-likelihood
    \begin{equation} \label{eqn:WVL.loglikelihood}   
        l_{ALP}(\boldsymbol{\gamma}) = \sum_{i \in B}\text{log}(p_i) + \sum_{i \in A} d^A_i \text{log}(1 - p_i)
    \end{equation}
    where $p_i = {\pi(\boldsymbol{x}_i; \boldsymbol{\theta})}/(1 + \pi(\boldsymbol{x}_i; \boldsymbol{\theta}))$ is the modelled probability of being in the sample $B$ given the unit is in the stacked sample.  \citet{wang.etal_2021} assume a logistic model for $p_i$, in other words $p_i = \text{exp}(\boldsymbol{x}^T_i\hat{\boldsymbol{\theta}})/(1 + \text{exp}(\boldsymbol{x}^T_i\hat{\boldsymbol{\theta}}))$.  This essentially means that the $\pi^B_i$ in the ALP method are implicitly modelled as $\pi(\boldsymbol{x}_i,\boldsymbol{\theta}) = \text{exp}(\boldsymbol{x}^T_i\boldsymbol{\theta})$, which contrasts with the other methods examined in this paper where the $\pi^B_i$ are generally assumed to follow a logistic model.
\end{remark}

\begin{remark}
    Note that when $B$ is a very large non-probability dataset, it becomes possible for the reference sample $A$ to have complete overlap with units in $B$.  In other words, the set $A \setminus B$ may be empty.  When this occurs, the OE, KW and SL approaches will not be able to be used as they require quantities to be calculated based on $A \setminus B$.  The CLW and ALP approaches are not affected by this limitation.
\end{remark}

Under a logistic regression model for the propensity scores, $\pi^B_i = \text{exp}(\boldsymbol{x}^T_i\boldsymbol{\theta}_0)/(1 + \text{exp}(\boldsymbol{x}^T_i\boldsymbol{\theta}_0))$.  Then the pseudo-likelihood function (\ref{eqn:new.loglikelihood}) becomes
\begin{align}
    l_{OE}(\boldsymbol{\theta}) 
    &= \sum_{i \in B} \text{log} \Bigl\{ \frac{\text{exp}({\boldsymbol{x}^T_i\boldsymbol{\theta}})}{1 + \text{exp}({\boldsymbol{x}^T_i\boldsymbol{\theta}})} \Bigr\} + \sum_{i \in A \setminus B}d^A_i \text{log} \Bigl\{ 1 - \frac{\text{exp}({\boldsymbol{x}^T_i\boldsymbol{\theta}})}{1 + \text{exp}({\boldsymbol{x}^T_i\boldsymbol{\theta}})}\Bigr\} \nonumber \\
    &= \sum_{i \in B} \Bigl\{ \boldsymbol{x}^T_i\boldsymbol{\theta} - \text{log}(1 + \text{exp}(\boldsymbol{x}^T_i\boldsymbol{\theta})) \Bigr\} - \sum_{i \in A \setminus B} d^A_i \text{log}(1 + \text{exp}(\boldsymbol{x}^T_i\boldsymbol{\theta})). \nonumber
\end{align}

Define the score function $S(\boldsymbol{\theta})$ as
\begin{align}
    S(\boldsymbol{\theta})
    &= \frac{\partial }{\partial \boldsymbol{\theta}}l_{OE}(\boldsymbol{\theta}) \nonumber \\
    &= \sum_{i \in B} \boldsymbol{x}_i - \sum_{i \in B}\pi(\boldsymbol{x}_i,\boldsymbol{\theta})\boldsymbol{x}_i - \sum_{i \in A \setminus B}d^A_i\pi(\boldsymbol{x}_i,\boldsymbol{\theta})\boldsymbol{x}_i \nonumber
\end{align}

We can obtain the maximum pseudo-likelihood estimator $\hat{\boldsymbol{\theta}}$ by solving $S(\boldsymbol{\theta}) = \boldsymbol{0}$.  A solution can be found by applying an iterative procedure such as Newton-Raphson
\begin{align}
    \boldsymbol{\theta}^{(m+1)} &=  \boldsymbol{\theta}^{(m)} + \Bigl\{ H(\boldsymbol{\theta}^{(m)}) \Bigr\}^{-1} S(\boldsymbol{\theta}^{(m)}) \nonumber
\end{align}

where
\begin{align} 
    H(\boldsymbol{\theta}^{(m)}) &= - \frac{\partial^2 }{\partial \boldsymbol{\theta}^2} l_{OE}(\boldsymbol{\theta}) \nonumber \\
    &=  \sum_{i \in B} \pi(\boldsymbol{x}_i,\boldsymbol{\theta})(1 - \pi(\boldsymbol{x}_i,\boldsymbol{\theta}))\boldsymbol{x}_i\boldsymbol{x}^T_i + \sum_{i \in A \setminus B}d^A_i\pi(\boldsymbol{x}_i,\boldsymbol{\theta})(1 - \pi(\boldsymbol{x}_i,\boldsymbol{\theta}))\boldsymbol{x}_i\boldsymbol{x}^T_i \nonumber
\end{align}

The OE participation probabilities for $i \in B$ are obtained from the maximum pseudo-likelihood estimates of $\hat{\boldsymbol{\theta}}$ through
\begin{align}
    \hat{\pi}^B_i = \frac{\text{exp}(\boldsymbol{x}^T_i\boldsymbol{\hat{\theta}})}{(1 + \text{exp}(\boldsymbol{x}^T_i\boldsymbol{\hat{\theta}}))} \nonumber
\end{align}

These participation probabilities can be fed into (\ref{eqn:IPW}) to produce an estimate of the population mean.  Alternatively, the participation probabilities may be used in a doubly robust manner such as described in \citet{chen.etal_2020}, or in multiply robust approaches such as those described in \citet{haziza.chen_2022}.

Table \ref{tab:methods.compare} provides a summary of the information requirements for the various methods described in this section.  For completeness, we have also included a row entry for the OE-EM approach, which is formally introduced in Section \ref{em_approach}.  The OE-EM method is a variation of the OE method that can be applied when the overlap between $A$ and $B$ is not known.

\begin{table}[ht]
    \centering
    \caption{Comparison of method information requirements}
      \begin{tabular}{lcccc}
        \hline
      Method & Overlap & $A \perp B$ required? & Treatment of & Quantity modelled \\
        &  known? &  & overlapping units &  \\
        \hline
        OE    & Yes  & No$^*$  & Excluded from $A$ & $\pi^B_i$ \\
        OE-EM & No   & Yes   & Estimated via EM  &  $\pi^B_i$ \\
        CLW   & No   & No   & Included in $A$ & $\pi^B_i$ \\
        KW    & Yes  & No$^*$   & Used to estimate for $B$ & $\pi^B_i$ \\
        ALP   & No   & No   & Stacked/duplicated  & $\pi^B_i/(1 + \pi^B_i)$ \\
        SL    & Yes  & Yes  & Excluded (no duplicates) & $\pi^B_i/(\pi^B_i + \pi^A_i - \pi^B_i\pi^A_i)$ \\
        \hline
      \end{tabular}
      \begin{minipage}{16cm}
			\vspace{0.1cm}
			\small $^*$Requires conditional independence, where knowledge of $\delta$ contains no further information about inclusion probabilities for $A$ once design variables $\boldsymbol{q}$ are taken into account.
		\end{minipage}
    \label{tab:methods.compare}
\end{table}

\subsection{Consistency and variance of the OE IPW estimator}

In order to examine the consistency and variance of the OE IPW estimator, we adopt an asymptotic framework used by other authors (see for example \citealt{chen.etal_2020}, \citealt{wang.etal_2021} and \citealt{scholtus.etal_2024}).  Within this framework, suppose there is a sequence of finite populations $\mathcal{U}_t$ of size $N_t$ indexed by $t$.  Associated with each $\mathcal{U}_t$ are a probability sample $A_t$ of size $n_{A,t}$ and non-probability sample $B_t$ of size $N_{B,t}$.  The population size $N_t \to \infty$, while the sample sizes $n_{A,t} \to \infty$ and $n_{B,t} \to \infty$ as $t \to \infty$.  We omit the index $t$ for the rest of this paper, and represent the limiting process by $N \to \infty$.  We describe the properties of the estimator under the model for the propensity scores and the probability sample design.

\begin{theorem} \label{thm:oe.var}
    When assumptions \textbf{A1} to \textbf{A3} from Section \ref{setup} and conditions \textbf{C1} to \textbf{C4} in Appendix \ref{sec:regularity_conditions} hold, $I_i$ and $\delta_i$ are independent for all $i$, and a logistic regression model for the participation probabilities holds, $E(\hat{\mu}_{OE}) \to \mu_y$ and $\lvert \hat{\mu}_{OE} - \mu_y \rvert \overset{p} \to 0$ as the sizes of the population and samples tend toward infinity.  Further, the variance of the OE IPW estimator, denoted by $Var(\hat{\mu}_{OE})$, is
\begin{align}
    Var(\hat{\mu}_{OE}) &\approx \frac{1}{N^2}\sum_{i \in \mathcal{U}} (1-\pi^B_i)\pi^B_i \Bigl[\frac{(y_i - \mu_y)}{(\pi^B_i)} - \boldsymbol{b}^T\boldsymbol{x}_i  \Bigr]^2 + \label{eqn:varIPW CLW term} \\
    & \boldsymbol{b}^T\boldsymbol{V}\boldsymbol{b} \text{ } +  \label{eqn:varIPW desvar} \\
    &  \frac{1}{N^2}\sum_{i \in \mathcal{U}}\Bigl[(\pi^B_i)^3(1 - \pi^B_i)(d^A_i - 1) \boldsymbol{b}^T\boldsymbol{x}_i\boldsymbol{x}^T_i\boldsymbol{b} \Bigr] \label{eqn:varIPW addterm}
\end{align}

where $\boldsymbol{b}^T = \sum_{i \in \mathcal{U}}(1 - \pi^B_i)(y_i - \mu)\boldsymbol{x}^T_i \Bigl[ \sum_{i \in \mathcal{U}}\pi^B_i(1 - \pi^B_i)\boldsymbol{x}_i\boldsymbol{x}^T_i \Bigr]^{-1}$, \\ $\boldsymbol{V} = \frac{1}{N^2} Var_d \Bigl(\sum_{i \in A}d^A_i\pi^B_i(1-\pi^B_i)\boldsymbol{x}_i \Bigr)$, and $Var_d$ refers to the design-based variance under the probability sample design for $A$.
\end{theorem}

Details of the derivations for consistency and variance can be found in Appendix \ref{sec:variance_derivation}.

A plug-in estimator can be constructed to estimate the variance of the OE IPW estimator (\ref{eqn:IPW}):
\begin{align}
    \widehat{Var}(\hat{\mu}_{OE})
    &= \frac{1}{N^2}\sum_{i \in B} (1-\hat{\pi}^B_i) \Bigl[\frac{(y_i - \hat{\mu}_{IPW})}{(\hat{\pi}^B_i)} - \boldsymbol{\hat{b}}^T\boldsymbol{x}_i  \Bigr]^2 + \label{eqn:varIPW CLW term plugin} \\
    & \boldsymbol{\hat{b}}^T\boldsymbol{\hat{V}}\boldsymbol{\hat{b}} \text{ } +  \label{eqn:varIPW desvar plugin} \\
    &  \frac{1}{N^2}\sum_{i \in B}\Bigl[(\hat{\pi}^B_i)^2(1 - \hat{\pi}^B_i)(d^A_i - 1) \boldsymbol{\hat{b}}^T\boldsymbol{x}_i\boldsymbol{x}^T_i\boldsymbol{\hat{b}} \Bigr] \label{eqn:varIPW addterm plugin}
\end{align}

where $\boldsymbol{\hat{b}}^T = \sum_{i \in B}(\frac{1}{\hat{\pi}^B_i} - 1)(y_i - \hat{\mu}_{IPW})\boldsymbol{x}^T_i \Bigl[ \sum_{i \in B}(1 - \hat{\pi}^B_i)\boldsymbol{x}_i\boldsymbol{x}^T_i \Bigr]^{-1}$, $\hat{\pi}^B_i$ is the estimated propensity for unit $i$ obtained using $\hat{\boldsymbol{\theta}}$ which is ``plugged-in'' for each unit $i$ as required in $A$ and $B$, $\hat{\mu}_{IPW}$ is the IPW estimate of mean for the variable of interest given by (\ref{eqn:IPW}), and $\boldsymbol{\hat{V}} = \frac{1}{N^2} \widehat{Var}_d \Bigl(\sum_{i \in A}d^A_i\hat{\pi}^B_i(1-\hat{\pi}^B_i)\boldsymbol{x}_i \Bigr)$ is the design-based variance for the quantity $\sum_{i \in \mathcal{U}} \hat{\pi}^B_i(1-\hat{\pi}^B_i)\boldsymbol{x}_i$ in $A$, estimated using standard methods.  When $B$ is small relative to $A$, then we may wish to use $\boldsymbol{\hat{b}}^T = \sum_{i \in B}(\frac{1}{\hat{\pi}^B_i} - 1)(y_i - \hat{\mu}_{IPW})\boldsymbol{x}^T_i \Bigl[ \sum_{i \in A}d^A_i\hat{\pi}^B_i (1 - \hat{\pi}^B_i)\boldsymbol{x}_i\boldsymbol{x}^T_i \Bigr]^{-1}$ instead (see \citealt{chen.etal_2020}).

Component (\ref{eqn:varIPW addterm plugin}) requires knowledge of the reference sample design weights of units in $B$, and this may not be available.  An alternative plug-in estimator for (\ref{eqn:varIPW addterm plugin}) is
\begin{align}
    \frac{1}{N^2}\sum_{i \in A} (d^A_i)^2\Bigl[(\hat{\pi}^B_i)^3(1 - \hat{\pi}^B_i)
    \hat{\boldsymbol{b}}^T\boldsymbol{x}_i\boldsymbol{x}^T_i\hat{\boldsymbol{b}} \Bigr] - 
    \frac{1}{N^2}\sum_{i \in B} \Bigl[(\hat{\pi}^B_i)^2(1 - \hat{\pi}^B_i)\hat{\boldsymbol{b}}^T\boldsymbol{x}_i\boldsymbol{x}^T_i\hat{\boldsymbol{b}} \Bigr]. \label{eqn:alt_plugin}
\end{align}
Again, if $B$ is small relative to $A$, then the second term of (\ref{eqn:alt_plugin}) may alternatively be estimated by $1/N^2\sum_{i \in A} d^A_i \Bigl[(\hat{\pi}^B_i)^3(1 - \hat{\pi}^B_i)\hat{\boldsymbol{b}}^T\boldsymbol{x}_i\boldsymbol{x}^T_i\hat{\boldsymbol{b}} \Bigr]$.

\subsection{Comparison of variance against alternative methods}

The OE pseudo-likelihood aims to take advantage of knowledge about $\delta_i, i \in A$.  As such, it may be expected to provide some variance gains over other methods which do not take advantage of this information.  In this subsection we compare the variance of the OE IPW estimator with the variances from the H\'ajek-like CLW2 estimator (denoted by $Var(\hat{\mu}_{CLW})$) in \citet{chen.etal_2020}, the data integration IPW estimator ($Var(\hat{\mu}_{KW})$) in \citet{kim.wang_2019}, the ALP method ($Var(\hat{\mu}_{ALP})$) in \citet{wang.etal_2021} and the SL approach proposed by \citet{beaumont.etal_2024b}.  To simplify our variance comparisons, we assume in this subsection that the reference sample $A$ and the non-probability sample $B$ have been drawn using Poisson sampling.  When this is the case, the variance of the estimator of population mean (\ref{eqn:IPW}) for each of the five methods can be expressed using
\begin{align}
    Var(\hat{\mu}) \approx \sum_{i \in \mathcal{U}} \Bigl[\pi^B_i(1 - \pi^B_i)\Bigl(\frac{y_i - \mu_y}{\pi^B_iN} - A_i \Bigr)^2 + \frac{1 - \pi^A_i}{\pi^A_i}(\pi^B_i)^2\lambda_iA^2_i \Bigr]. \label{eqn:commonvar_all}
\end{align}
Table \ref{tab:commonvar_terms} provides the terms for each of the different methods.  See the Supplementary Materials for details on the derivation of the variance expressions.

\begin{table}[ht]
    \centering
    \caption{Variance factors for different methods}
      \begin{tabular}{lccc}
        \hline
        Method & $A_i$ & $\lambda_i$ & $\boldsymbol{G}$\\
        \hline
        \addlinespace[0.8ex]
        OE & $\boldsymbol{c}^T\boldsymbol{G}^{-1}\dfrac{\pi'^B_i}{\pi^B_i(1 - \pi^B_i)}$ & $1 - \pi^B_i$ & $\sum_{i \in \mathcal{U}} \dfrac{\pi'^B_i(\pi'^B_i)^T}{\pi^B_i(1 - \pi^B_i)}$ \\ [2.4ex]
        CLW & $\boldsymbol{c}^T\boldsymbol{G}^{-1}\dfrac{\pi'^B_i}{\pi^B_i(1 - \pi^B_i)}$ & 1 & $\sum_{i \in \mathcal{U}} \dfrac{\pi'^B_i(\pi'^B_i)^T}{\pi^B_i(1 - \pi^B_i)}$ \\ [2.4ex]
        KW & $\boldsymbol{c}^T\boldsymbol{G}^{-1}\dfrac{\pi'^B_i}{\pi^B_i(1 - \pi^B_i)}$ & $\dfrac{1 - \pi^B_i}{\pi^B_i}$ & $\sum_{i \in \mathcal{U}} \dfrac{\pi'^B_i(\pi'^B_i)^T}{\pi^B_i(1 - \pi^B_i)}$ \\ [2.4ex]
        ALP & $\boldsymbol{c}^T\boldsymbol{G}^{-1}\dfrac{\pi'^B_i}{\pi^B_i(1 + \pi^B_i)}$ & 1 & $\sum_{i \in \mathcal{U}} \dfrac{\pi'^B_i(\pi'^B_i)^T}{\pi^B_i(1 + \pi^B_i)}$ \\ [2.4ex]
        SL & $\boldsymbol{c}^T\boldsymbol{G}^{-1}\dfrac{\phi_i \pi'^B_i}{\pi^B_i(1 - \pi^B_i)}$ & $1 - \pi^B_i$ & $\sum_{i \in \mathcal{U}} \dfrac{\phi_i\pi'^B_i(\pi'^B_i)^T}{\pi^B_i(1 - \pi^B_i)}$ \\ [2.4ex]
        \hline
      \end{tabular}

      \begin{minipage}{11cm}
			\vspace{0.1cm}
			\small Note: $\boldsymbol{c}^T$ is defined as $1/N\sum_{i \in \mathcal{U}}(\pi'^B_i)^T/\pi^B_i \Bigl( y_i - \mu_y \Bigr)$, $\phi_i$ is defined as $\pi^A_i/(\pi^B_i + \pi^A_i - \pi^B_i\pi^A_i)$, and $\pi'^B_i$ is the first derivative of $\pi(\boldsymbol{x}_i;\boldsymbol{\theta})$ evaluated at $\boldsymbol{\theta}_0$.
	    \end{minipage}
        
    \label{tab:commonvar_terms}
  \end{table}

From Table \ref{tab:commonvar_terms}, we can see that the $A_i$ and $\boldsymbol{G}$ terms for the OE, CLW and KW methods are the same.  The variance expressions only vary through the $\lambda_i$ factor which is in the second term of (\ref{eqn:commonvar_all}).  It is straightforward to see that for any value of $\pi^B_i$ the OE IPW estimator is thus always at least as efficient as the CLW and KW estimators when the reference sample is a Poisson sample.  Its efficiency over the CLW approach increases when $\{\pi^A_i\}$ are smaller, and when $\{\pi^B_i\}$ are larger.  Its efficiency over the KW data integration estimator increases when $\{\pi^A_i\}$ are smaller, or when $\{\pi^B_i\}$ are smaller.  The KW estimator becomes more efficient than the CLW estimator when $\{\pi^B_i\}$ are larger (in fact, when $\pi^B_i$ tend to be greater than 0.5).

In comparing the OE and ALP variance terms in Table \ref{tab:commonvar_terms}, we note firstly that the $A_i$ term is a weighted least squares fit for $(y_i-\mu_y)/\pi^B_iN$.  To see this, let $\boldsymbol{\beta}=\boldsymbol{G}^{-1}\boldsymbol{c}$, so $\boldsymbol{\beta}$ solves the equation $\boldsymbol{G\beta} = \boldsymbol{c}$.  For each of the five methods, this can be expressed as
\begin{align}
    &\sum_{i \in \mathcal{U}}\frac{\pi'^B_i(y_i - \mu_y)}{\pi^B_iN}  - \sum_{i \in \mathcal{U}} \pi'^B_i\boldsymbol{v}^T_i\boldsymbol{\beta} = 0 \nonumber \\ 
    \implies & \sum_{i \in \mathcal{U}}w_i\boldsymbol{v}_i\Bigl(\frac{y_i-\mu_y}{\pi^B_iN} - \boldsymbol{v}^T_i\boldsymbol{\beta} \Bigr) = 0 \label{eqn:wls_equation}
\end{align}
which is in the form of a weighted least squares normal equation where $\boldsymbol{\beta}^T\boldsymbol{v}_i$ predicts $(y_i - \mu_y)/(\pi^BN)$, with weights $w_i$.  For the OE approach, $\boldsymbol{v}_i=\pi'^B_i/[\pi^B_i(1 - \pi^B_i)]$ and $w_i = \pi^B_i(1 - \pi^B_i)$, while for the ALP approach $\boldsymbol{v}_i=\pi'^B_i/[\pi^B_i(1 + \pi^B_i)]$ and $w_i = \pi^B_i(1 + \pi^B_i)$.

When the auxiliary variables $\boldsymbol{x}$ explain a large part of the variation in $(y_i-\mu_y)/(\pi^B_iN$), and $A$ is small relative to $B$, the second variance term in (\ref{eqn:commonvar_all}) will tend to be dominant. Given that the $\lambda_i$ factor for OE is $1 - \pi^B_i$ while it takes a value of 1 for the ALP approach, the second term will tend to be smaller for the OE approach when the $\pi^B_i$ values are larger, or in other words, when $B$ consists of a large proportion of the population.

Comparing the OE and SL variance terms in Table \ref{tab:commonvar_terms}, the $\phi_i$ factor present in $A_i$ and $\boldsymbol{G}$ is the only difference.  The two methods achieve the same variance when $\phi_i=1$ or when $\phi_i$ is constant.  The first case will occur when $A$ is a census.  The second case would be satisfied when the homogeneous group model described in \citet{beaumont.etal_2024b} applies, and $\pi^A_i$ are constant within each group.  In the homogeneous group model, the auxiliary variables partition the population into groups with $\pi^B_i$ being constant in each group, so $\pi^A_i$ would be constant in each group when $A$ is a simple random sample, or a stratified simple random sample where groups are then formed using the cross-classification of stratum variables and the auxiliary variable groups.

Further, the variance for both the OE and SL approaches can be expressed in the form
\begin{align}
    Var(\hat{\mu}) &= \sum_{i \in \mathcal{U}} \pi^B_i(1 - \pi^B_i) \Bigl[ \gamma_i\pi^B_iA^2_i + (C_i - A_i)^2 \Bigr] \nonumber \\
    &= \sum_{i \in \mathcal{U}} \pi^B_i(1 - \pi^B_i) \Bigl[ (1 + \gamma_i\pi^B_i)A^2_i + C^2_i - 2C_iA_i\Bigr] \nonumber \\
    &= \sum_{i \in \mathcal{U}} \Bigl[ \frac{1}{\phi_i}\Bigl(A_i - \phi_iC_i \Bigr)^2 + (1 - \phi_i)C^2_i \Bigr] \label{eqn:SL_OE_var}
\end{align}
where $\gamma_i = (1 - \pi^A_i)/\pi^A_i$, $C_i = (y_i-\mu_y)/(\pi^B_iN)$, and $A_i$ and $\phi_i$ are defined in Table \ref{tab:commonvar_terms}.  The second variance term is the same for both approaches, so only the first term is of interest for our comparison.  This term will be minimised when $A_i = \phi_iC_i$, but we know from (\ref{eqn:wls_equation}) that both approaches have $A_i$ terms which are predicting or fitting the quantity $C_i$.  Note that in the normal equation (\ref{eqn:wls_equation}) for the SL approach we have a $\phi_i$ factor multiplying the $\boldsymbol{v}^T_i\boldsymbol{\beta}$ term, so the SL fitted values will not achieve as good a fit as the OE approach.

Let $\boldsymbol{\beta}_S = \boldsymbol{G}^{-1}_S\boldsymbol{c}$ where $\boldsymbol{G}^{-1}_S$ is the $\boldsymbol{G}$ matrix for the SL approach, and in similar fashion $\boldsymbol{\beta}_O = \boldsymbol{G}^{-1}_O\boldsymbol{c}$.  Then, with $\boldsymbol{v}_i=\pi'^B_i/[\pi^B_i(1 - \pi^B_i)]$ we can see that
\begin{align}
    \frac{A^{OE}_i}{A^{opt}_i} = \frac{\boldsymbol{\beta}^T_O\boldsymbol{v}_i}{\phi_iC_i}, \text{ and }\frac{A^{SL}_i}{A^{opt}_i} = \frac{\phi_i\boldsymbol{\beta}^T_S\boldsymbol{v}_i}{\phi_iC_i} = \frac{\boldsymbol{\beta}^T_S\boldsymbol{v}_i}{C_i}. \nonumber
\end{align}
When the auxiliary variables $\boldsymbol{x}$ enable a good fit, the ratio $\boldsymbol{\beta}^T_O\boldsymbol{v}_i/C_i$ for the OE approach is approximately 1, and so $A^{OE}_i/A^{opt}_i \approx 1/\phi_i$.  The ratio $\boldsymbol{\beta}^T_S\boldsymbol{v}_i/C_i$ will be larger relative to the OE approach, but $A^{SL}_i/A^{opt}_i$ should tend to be closer to 1 since it does not have the $1/\phi_i$ factor.  The SL approach would thus tend to achieve a lower variance than the OE approach in this case.

There will be a point where the poorer ability of the SL approach to predict $C_i$ will eventually make $\boldsymbol{\beta}^T_S\boldsymbol{v}_i/C_i$ much larger than $\boldsymbol{\beta}^T_O\boldsymbol{v}_i/C_i$, so that it overcomes the $1/\phi_i$ penalty in the OE ratio.  This will occur when the $\boldsymbol{x}$ variables enable only a moderate to poor fit for $C_i$, and the $\phi_i$ factors tend to vary only moderately or are large.

Note that the SL approach also requires more restrictive assumptions to hold.  In particular, it is assumed that the samples $A$ and $B$ are mutually independent so that $P(i \in A \cap B) = P(i \in A)P(i \in B) = \pi^A_i\pi^B_i$.  In practice, this may be achieved by ensuring that there is independence in the sample design and the selection mechanism (for example if both $A$ and $B$ are drawn as independent Poisson samples), or when the inclusion probabilities into the two samples do not depend on shared (observed or unobserved) information.  We explore these assumptions further in Section \ref{real_data}.

\begin{remark}
    There is equivalence between $Var(\hat{\mu}_{CLW})$, $Var(\hat{\mu}_{KW})$, $Var(\hat{\mu}_{SL})$ and $Var(\hat{\mu}_{OE})$ when $\pi^A_i=1$, that is, when the reference sample $A$ is in fact the full population and $\boldsymbol{x}_i$ is available for all $i \in \mathcal{U}$.  In this case, knowledge about $\delta_i, i \in \mathcal{U}$ does not give any advantages for the SL, OE and \citet{kim.wang_2019} approaches over the \citet{chen.etal_2020} approach.
\end{remark}

\section{OE propensity score estimation when data integration is not possible} \label{em_approach}

The proposed OE pseudo log-likelihood (\ref{eqn:new.loglikelihood}) requires knowledge of $\delta_i$ for each unit in $A$.  In some situations, knowledge of $\delta_i, i \in A$ will not be available.  To address unknown overlap when implementing their SL approach, \citet{beaumont.etal_2024b} proposed to estimate the term involving the overlapping sample in their likelihood function (\ref{eqn:SL.loglikelihood}) using a best linear unbiased prediction approach.  In this section we describe an alternative approach to the problem by treating the unknown $\delta_i$ as a missing-data problem (see \citealt{kim.tam_2021}).  A two-step algorithm with similarities to the EM algorithm of \citealt{dempster.etal_1977} is employed to simultaneously estimate the missing $\delta_i$ values and obtain the maximum likelihood estimator for (\ref{eqn:new.loglikelihood}) (see for example Chapter 3.3 in \cite{kim.shao_2021}).  We first re-write the OE pseudo-loglikelihood (\ref{eqn:new.loglikelihood}) as
\begin{align}
    l_{OE}(\boldsymbol{\theta}) &= \sum_{i \in B} \text{log}\pi(\boldsymbol{x}_i;\boldsymbol{\theta}) + \sum_{i \in A} d^A_i\text{log}(1 - \pi(\boldsymbol{x}_i;\boldsymbol{\theta})) - \sum_{i \in A}d^A_i\delta_i\text{log}(1 - \pi(\boldsymbol{x}_i;\boldsymbol{\theta})). \label{eqn:em ll}
\end{align}

Under a logistic regression model for $\pi^B_i$, (\ref{eqn:em ll}) becomes
\begin{align}
    l_{OE}(\boldsymbol{\theta})
    &= \sum_{i \in B} \Bigl\{ \boldsymbol{x}^T_i\theta - \text{log}(1 + \text{exp}(\boldsymbol{x}^T_i\theta)) \Bigr\} - \sum_{i \in A} d^A_i \text{log}(1 + \text{exp}(\boldsymbol{x}^T_i\theta)) + \nonumber \\
    &\mathrel{\phantom{=}} \sum_{i \in A} d^A_i \delta_i \text{log}(1 + \text{exp}(\boldsymbol{x}^T_i\theta)) \nonumber
\end{align}

and the score function $S(\boldsymbol{\theta})$ is given as
\begin{align}
    S(\boldsymbol{\theta})
    &= \frac{\partial }{\partial \boldsymbol{\theta}}l_{OE}(\boldsymbol{\theta}) \nonumber \\
    &= \sum_{i \in B} \boldsymbol{x}_i - \sum_{i \in B}\pi(\boldsymbol{x}_i,\boldsymbol{\theta})\boldsymbol{x}_i - \sum_{i \in A}d^A_i\pi(\boldsymbol{x}_i,\boldsymbol{\theta})\boldsymbol{x}_i + \sum_{i \in A}d^A_i \delta_i \pi(\boldsymbol{x}_i,\boldsymbol{\theta})\boldsymbol{x}_i. \nonumber
\end{align}

The EM-like procedure proceeds as follows:

\textit{Step 1:}
Calculate initial values of the parameters, $\boldsymbol{\theta}^{(1)}$, based on the information available.  These can be obtained using a method like \citet{chen.etal_2020} which does not require knowledge of $\delta_i$.  The values $\boldsymbol{\theta}^{(1)}$ are used to obtain $\pi^{B,(1)}_i$.

\textit{Step 2:}
The Expectation step assumes the two samples $A$ and $B$ are independent.  We calculate the expected value of $l_{OE}(\boldsymbol{\theta})$ with respect to $\delta_i \sim P(\cdot | \boldsymbol{x}, \boldsymbol{\theta}^{(j)})$, holding the sample $A$ as fixed.  The expectation is given by
\begin{align}
    E_{\delta_i \sim P(\cdot | \boldsymbol{x},  \boldsymbol{\theta}^{(j)})}[l_{OE}(\boldsymbol{\theta})]
    &= \sum_{i \in B} \Bigl\{ \boldsymbol{x}^T_i\boldsymbol{\theta} - \text{log}(1 + \text{exp}(\boldsymbol{x}^T_i\boldsymbol{\theta})) \Bigr\} - \sum_{i \in A} d^A_i \text{log}(1 + \text{exp}(\boldsymbol{x}^T_i\boldsymbol{\theta})) \nonumber \\
    & + \sum_{i \in A} d^A_i E(\delta_i|\boldsymbol{x},\boldsymbol{\theta}^{(j)}) \text{log}(1 + \text{exp}(\boldsymbol{x}^T_i\boldsymbol{\theta})). \label{eqn:e-step}
\end{align}

Note that $E(\delta_i|\boldsymbol{x}, \boldsymbol{\theta}^{(j)})$ can be interpreted as the conditional probability of unit $i$ being included in the non-probability sample $B$, given the auxiliary variables $\boldsymbol{x}$ and $\boldsymbol{\theta}^{(j)}$.  Under a logistic model we estimate this expectation for each unit using
\begin{align*}
    E(\delta_i|\boldsymbol{x}, \boldsymbol{\theta}^{(j)}) = \hat{\delta}^{(j)}_i &= \pi^{B,(j)}_i = \frac{\text{exp}(\boldsymbol{x}^T_i\boldsymbol{\theta}^{(j)})}{(1 + \text{exp}(\boldsymbol{x}^T_i\boldsymbol{\theta}^{(j)})}
\end{align*}
where $\pi^{B,(j)}_i$ is the value of $\pi^B_i$ for the $j$'th iteration.

\textit{Step 3:}
In the Maximisation step, the log-likelihood (\ref{eqn:e-step}) is maximised with respect to $\boldsymbol{\theta}$ to obtain $\boldsymbol{\theta}^{(j+1)}$.  The maximisation can generally be accomplished using an iterative procedure such as the Newton-Raphson method described in Section \ref{oe_propscores}.

\textit{Step 4:}
Set $j = j + 1$, and go back to Step 2.  Repeat until convergence - that is, $\text{max}(\lvert \boldsymbol{\theta}^{(j)} - \boldsymbol{\theta}^{(j-1)} \rvert ) < \epsilon$, where $\text{max}(\boldsymbol{x})$ returns the component from vector $\boldsymbol{x}$ that is the largest, and $\epsilon$ is a user-specified threshold.

The final set of parameters $\hat{\boldsymbol{\theta}}$ are used to obtain $\hat{\pi}^B_i$ and subsequently may be fed into (\ref{eqn:IPW}) to produce the IPW estimate of population mean.

\begin{remark}
    In the special case where $A$ is a Census, the OE, KW and SL approaches reduce to the full likelihood estimate like CLW.  From empirical testing, the proposed EM-like procedure does not appear to converge exactly to the full likelihood estimate.  However, it was found to produce results very close to the full likelihood estimate, with parameter estimates for $\hat{\boldsymbol{\theta}}$ generally equal to those from the CLW approach up to the second decimal place, and with relative bias and relative root mean square error results essentially equivalent to those obtained from the CLW approach.
\end{remark}

\begin{algorithm}[ht]
    \caption{Bootstrap variance estimation for EM algorithm}\label{alg:em.bootstrap}
    \begin{algorithmic}[1]
        \Require $A, B, n_A, n_B, b_{rep}$ \;
        \State Set $r := 0$ \; 
        \For{$r = 1, ...,b_{rep}$}
            \State Select a with-replacement sample $A_{r}$ from $A$ with size $n_A$
            \State Select a with-replacement sample $B_r$ from $B$ with size $n_B$
            \State Calculate $\hat{\boldsymbol{\theta}}^r$ using $A_r$ and $B_r$ based on the EM-like algorithm, where the weight used for unit $i$ is its original weight, $d^A_i$
            \State Calculate $\hat{\pi}^{B_r}_i$ for each unit $i \in B_r$ using $\hat{\boldsymbol{\theta}}^r$
            \State Calculate $\hat{Y}^r$ based on $\hat{\pi}^{B_r}_i$ and $B_r$
        \EndFor
        \State Return the variance of the bootstrap estimates $\hat{Y}^r, \text{where } r=\{1,..,b_{rep}\}$
    \end{algorithmic}
\end{algorithm}

Variance estimation for the IPW estimator which uses $\hat{\pi}^B_i$ from the EM-like algorithm may be undertaken using the bootstrap method.  Under a single-stage PPS sample design for $A$ with small sampling fractions, and similarly when $B$ is not large, we can use a with-replacement bootstrap (see \cite{rao.etal_1992} and \cite{chen.etal_2020}).  We present the bootstrap procedure in Algorithm \ref{alg:em.bootstrap}, where $b_{rep}$ is the number of bootstrap replications used.  Note that using with-replacement sampling for $B$ in the algorithm may not be so appropriate when $B$ is large and has a non-negligible sampling fraction.  In this case, employing a bootstrap procedure which assumes that $B$ has been selected using Poisson sampling may be better.  Options include the approach described in \citet{berg.etal_2025} and originally proposed in \citet{antal.tille_2011}, or a pseudo-population bootstrap approach similar to that proposed in \citet{liu.etal_2022}.

\section{Simulation Study} \label{simstudy}

A simulation study was undertaken to examine the consistency of our proposed OE propensity scores, and to compare its efficiency with alternative methods.  The setup of the simulation is similar to what was used in \citet{chen.etal_2020}.  We consider a finite population of size $N = 200,000$ that includes a data item of interest $y$ and auxiliary variables $\boldsymbol{x}$.  The data items $y$ and $\boldsymbol{x}$ are related by the model
\begin{equation}
    y_i = \boldsymbol{x}^T_i \boldsymbol{\beta} + \sigma\epsilon_i \nonumber
\end{equation}

where $\boldsymbol{x}_i = (1, x_{1i}, x_{2i}, x_{3i}, x_{4i})$, $\boldsymbol{\beta}=(2,1,1,1,1)$, $x_{1i} = z_{1i}, x_{2i} = z_{2i} + 0.3x_{1i}, x_{3i} = z_{3i} + 0.2(x_{1i} + x_{2i}), x_{4i} = z_{4i} + 0.1(x_{1i} + x_{2i} + x_{3i})$, and $z_{1i} \sim \text{Bernoulli}(0.5), z_{2i} \sim \text{Uniform}(0,2), z_{3i} \sim \text{Exponential}(1)$, and $z_{4i} \sim \chi^2(4)$.  The error terms $\epsilon_i$ are independent and identically distributed as $N(0,1)$.  In our simulations, we utilise two values of $\sigma$, chosen such that the correlation coefficient $\rho$ between $y$ and $\boldsymbol{x}^T\boldsymbol{\beta}$ is either 0.3 (moderately weak) or 0.7 (moderately strong).

The true propensity scores $\pi^B_i$ are given by
\begin{equation}
    \text{log}\Bigl(\frac{\pi^B_i}{1 - \pi^B_i} \Bigr) = \theta_0 + 0.1x_{1i} + 0.2x_{2i} + 0.3x_{3i} + 0.4x_{4i} \nonumber
\end{equation}

where the intercept $\theta_0$ was chosen to ensure the target sample size for $N_B$ is achieved.  Poisson sampling was used to select the sample $B$, while randomised systematic probability proportional to size sampling (PPS) was used to select the reference sample $A$.  In general, the inclusion probabilities for $\pi^A_i$ were proportional to $z_i = c + x_{3i}$, where $c$ was chosen to ensure the ratio max($z_i$)/min($z_i$) was equal to 10.

The simulation exercises aimed to examine the performance of the OE IPW estimators relative to a number of other approaches, and to assess the effectiveness of the proposed variance estimators.  For the comparison of IPW estimators and the plug-in variance estimator, we produced results using two different sample sizes for $A$: a small sample of $n_A = 500$, and a more standard sample size of $n_A = 5,000$.  We produced results for several different sample sizes for $N_B$: 750, 2,000, 50,000 and 100,000 (for the smaller $A$ sample, to avoid occasional issues with singular matrices when applying the CLW method) or 140,000 (for the larger $A$ sample).  To assess the bootstrap estimator proposed in Algorithm \ref{alg:em.bootstrap} for the EM algorithm approach, we used the following sample sizes for $N_B$ - 750, 2,000, 50,000 and 90,000.  The use of the smaller maximum sample size of 90,000 was to enable the simulation to conclude in a reasonable amount of time.

$R = 2,000$ simulations were undertaken for each combination of $\rho$ and sample size.  For the EM algorithm approach, 200 bootstrap iterations were used to calculate the bootstrap variances.  The performance of the different estimators was evaluated by calculating the Monte Carlo Relative Bias (RB) and Relative Root Mean Squared Error (RRMSE):
\begin{align}
    \text{RB} = \frac{1}{R}\sum_{r=1}^{R}\frac{\hat{\bar{Y}}_r - \bar{Y}}{\bar{Y}}, \text{  RRMSE}=\frac{\sqrt{\frac{1}{R}\sum_{r=1}^{R}(\hat{\bar{Y}}_r-\bar{Y})^2}}{\bar{Y}} \nonumber
\end{align}

where $\hat{\bar{Y}}_r$ is the estimate of the population mean computed for the $r$'th simulation run and $\bar{Y}=\mu_Y$ is the true population mean.

Results were produced for the following estimators:
\begin{itemize}
    \item Naive - Simple sample mean from $B$
    \item OE - The OE IPW estimator
    \item OE-EM - The IPW estimator produced by applying the EM algorithm to estimate $\hat{\delta}_i, i \in A$ in the OE pseudo-likelihood
    \item SL - The sample likelihood approach described in \citet{beaumont.etal_2024b}
    \item CLW - The IPW2 estimator proposed by \citet{chen.etal_2020}
    \item ALP - The ALP estimator proposed by \citet{wang.etal_2021}
    \item KW - The estimator proposed by \citet{kim.wang_2019}
\end{itemize}
Note that results for the KW approach were not produced for the smaller $n_A=500$ sample size, since the approach relies on there being some overlapping sample between $A$ and $B$.  Smaller $A$ sample sizes would occasionally result in no overlapping sample.

A logistic model was used for $\pi^B_i$ when implementing most of the methods.  The exception was the ALP method where we assumed a logistic model for $p_i = \pi(\boldsymbol{x}_i, \boldsymbol{\theta})/(1 + \pi(\boldsymbol{x}_i, \boldsymbol{\theta}))$, to remain consistent with the approach described in \citet{wang.etal_2021}.

\subsection{Simulation Results}

Tables \ref{tab:sim1_results} and \ref{tab:sim1_results_small} compare RB and RRMSE values for each of the estimators under the different sample sizes $N_B$, for each of the two $n_A$ scenarios.  Results are shown for the case when the correlation between $y$ and $\boldsymbol{x}^T\beta$ was $\rho = 0.3$.  The results for $\rho = 0.7$ are similar to those when $\rho = 0.3$,and are provided in Appendix \ref{app:additional_results} for the interested reader.

The Naive estimate based solely on the sample mean of $B$ is very biased for all $N_B$ sizes, as expected given the selection bias present in $B$.  When the reference sample is larger, the proposed OE and OE-EM estimators, along with the CLW, KW and SL estimators, are all approximately unbiased.  There is a hint of bias in the ALP estimator for the smaller $N_B$ sample sizes, but the bias looks to dissipate as $N_B$ gets larger.

\begin{table}[ht]
    \centering
    \caption{Comparative performance of estimators - $n_A = 5,000$ and $\rho(y,\boldsymbol{x}^T\beta)=0.3$}
    \resizebox{0.995\textwidth}{!}{
      \begin{tabular}{lrrrrrrrr}
        \hline
      \multicolumn{1}{c}{} & \multicolumn{2}{c}{$N_B=750$} & \multicolumn{2}{c}{$N_B=2,000$}
        & \multicolumn{2}{c}{$N_B=50,000$} & \multicolumn{2}{c}{$N_B=140,000$} \\
       & \%RB & \%RRMSE & \%RB & \%RRMSE & \%RB & \%RRMSE & \%RB & \%RRMSE \\
        \hline
        Naive & 85.09 & 85.20 & 73.52 & 73.57 & 27.55 & 27.56 & 8.11  & 8.11 \\
        OE    & -0.13 & 9.84  & -0.06 & 5.80  & -0.01 & 0.77  & -0.02 & 0.32 \\
        OE-EM & 0.00  & 9.77  & 0.03  & 5.81  & 0.01  & 0.82  & 0.00  & 0.42 \\
        SL    & 0.09  & 9.63  & 0.11  & 5.63  & 0.01  & 0.71  & -0.01 & 0.31 \\
        CLW   & -0.33 & 10.21 & -0.17 & 6.16  & 0.25  & 1.15  & -0.06 & 0.74 \\
        ALP   & 0.89  & 9.47  & 1.10  & 5.59  & 0.47  & 0.88  & 0.04  & 0.53 \\
        KW    & 0.27  & 12.56 & 0.01  & 6.88  & -0.04 & 1.13  & -0.04 & 0.42 \\
      \hline
      \end{tabular}
      }
        \begin{minipage}{16cm}
			\vspace{0.1cm}
			\small Note: ``Naive'' - Simple sample mean from $B$; ``OE'' - Overlap-Excluded IPW estimator; ``OE-EM'' - Use of EM algorithm to estimate links between $A$ and $B$; ``SL'' - Sample likelihood approach; ``CLW'' - IPW2 estimator in \citet{chen.etal_2020}; ``ALP'' - ALP estimator in \citet{wang.etal_2021}; ``KW'' - Estimator proposed by \citet{kim.wang_2019}.
		\end{minipage}
    \label{tab:sim1_results}
  \end{table}

Under the scenario where the reference sample is small, the CLW, OE, and OE-EM methods exhibited a small bias when $B$ was also small.  This bias vanished for the OE and OE-EM approaches as $B$ increased.  While the bias also reduced for the CLW approach as $B$ increased, there was still a hint of bias remaining even at the largest sample size $N_B=100,000$.  On inspection, the bias appears to stem from a small reference sample size leading to a small number of units having very large design weights.  These can then have an adverse effect on the estimation of the $\boldsymbol{\theta}$ parameters.  The SL approach, which doesn't utilise the design weights from $A$, is unaffected by this issue.  The ALP method also appears to be less affected by bias in this scenario relative to when $n_A$ is large.

\begin{table}[ht]
    \centering
    \caption{Comparative performance of estimators - $n_A = 500$ and $\rho(y,\boldsymbol{x}^T\beta)=0.3$}
    \resizebox{0.995\textwidth}{!}{
      \begin{tabular}{lrrrrrrrr}
        \hline
      \multicolumn{1}{c}{} & \multicolumn{2}{c}{$N_B=750$} & \multicolumn{2}{c}{$N_B=2,000$}
        & \multicolumn{2}{c}{$N_B=50,000$} & \multicolumn{2}{c}{$N_B=100,000$} \\
       & \%RB & \%RRMSE & \%RB & \%RRMSE & \%RB & \%RRMSE & \%RB & \%RRMSE \\
        \hline
        Naive & 85.09 & 85.20 & 73.52 & 73.57 & 27.55 & 27.56 & 14.93 & 14.93 \\
        OE & -2.23 & 11.79 & -1.12 & 7.31  & -0.13 & 1.65  & -0.06 & 1.11 \\
        OE-EM & -1.01 & 11.71 & -0.39 & 7.16  & -0.04 & 1.88  & 0.02  & 1.43 \\
        SL & 0.12  & 9.62  & 0.15  & 5.68  & 0.03  & 1.28  & 0.00  & 0.99 \\
        CLW & -4.75 & 16.45 & -2.57 & 10.20 & -0.66 & 3.98  & -0.70 & 3.61 \\
        ALP & 0.41  & 9.58  & 0.47  & 5.65  & 0.12  & 1.72  & 0.06  & 1.63 \\
      \hline
      \end{tabular}
      }
        \begin{minipage}{16cm}
			\vspace{0.1cm}
			\small Note: ``Naive'' - Simple sample mean from $B$; ``OE'' - Overlap-Excluded IPW estimator; ``OE-EM'' - Use of EM algorithm to estimate links between $A$ and $B$; ``SL'' - Sample likelihood approach; ``CLW'' - IPW2 estimator in \citet{chen.etal_2020}; ``ALP'' - ALP estimator in \citet{wang.etal_2021}.
		\end{minipage}
    \label{tab:sim1_results_small}
  \end{table}

In terms of the RRMSE performance of the estimators, when $\rho=0.3$ the ALP estimator provided the most efficient estimates for the smallest $N_B$ sizes of 750 and 2,000.  When $N_B$ was 50,000 or 140,000 the SL estimator provided the lowest RRMSE results.  The proposed OE and OE-EM estimators tended not to perform as well as the SL and ALP estimators for the smaller $N_B$ sizes, but provided competitive RRMSE results when $N_B$ was larger.  The performance advantage of the SL approach over other methods was greater for the smaller reference sample size.  Though not shown here, the RRMSE results for $\rho=0.7$ are similar, however we note that for this higher correlation value the SL estimator provided the lowest RRMSE values even for the lowest $N_B$ values.

The KW estimator did not perform as well compared with most of the other estimators when $N_B$ was small.  Very small values of $N_B$, coupled with generally small to moderate reference sample sizes $n_A$, may result in a very small number of units in the domain $A \cap B$, and in the extreme case the domain may be empty.  Under larger values of $N_B$, the performance of the KW estimator improved.

The results from Tables \ref{tab:sim1_results} and \ref{tab:sim1_results_small} also show that estimating the values of $\hat{\delta}_i$ in $A$ using the EM algorithm produces consistent estimates of the population mean.  In the simulations, when $N_B$ was larger, the RRMSE tended to be slightly higher compared to when we have $\delta_i$ available.  However, the OE-EM RRMSE values were lower than those for the OE approach for the smaller $N_B$ values.

\begin{table}[ht]
    \centering
    \caption{Average overlap between $A$ and $B$ in simulations}
      \begin{tabular}{lrrrr}
        \hline
        Sample size of $A$ & $N_B = 750$ & $N_B = 2,000$ & $N_B = 50,000$ & $N_B = 140,000$ \\
        \hline
        500 & 2.2 & 5.8 & 135.7 & 263.0 \\
        5,000 & 21.6 & 57.5 & 1,357.2 & 3,608.1 \\
        \hline
      \end{tabular}
    \label{tab:average_overlap}
  \end{table}

Table \ref{tab:average_overlap} provides the average sample overlap between $A$ and $B$ for the Monte Carlo simulations under the various sample sizes for $N_B$, for each of the two reference sample sizes.

\begin{table}[ht]
    \centering
    \caption{Coverage Probabilities and Average Length of 95\% Confidence Intervals}
      \begin{tabular}{lrrrr}
        \hline
      \multicolumn{1}{c}{} & \multicolumn{2}{c}{$\rho(Y,X\beta)=0.3$} & \multicolumn{2}{c}{$\rho(Y,X\beta)=0.7$} \\
      \multicolumn{1}{c}{$N_B$} & Coverage Probability & Average Length & Coverage Probability & Average Length \\
        \hline
        750     &   0.945   &   3.71    &   0.951   &  1.66 \\
        2,000   &   0.940    &   2.12	&   0.956	&  0.99 \\
        50,000  &   0.955	&   0.29	&   0.952	&  0.19 \\
        140,000 &   0.951	&   0.11	&   0.946	&  0.09 \\
        \hline
      \end{tabular}
    \label{tab:coverage_rate}
\end{table}

We next examined the coverage probabilities (CP) and average length (AL) of the 95\% confidence intervals for the OE plug-in variance estimator, and the OE-EM bootstrap variance approach.  In this comparison, the design-based variance component (\ref{eqn:varIPW desvar plugin}) was estimated by assuming $A$ was drawn using with-replacement sampling.  This is a reasonable approximation to make for a without-replacement PPS sample when $N$ is large and the sampling fraction is small (\citealt[pp. 80-81]{wu.thompson_2020}).

Table \ref{tab:coverage_rate} provides the CP and AL values for the 95\% confidence intervals using the plug-in variance estimator for the OE IPW estimator.  Results are shown for both $\rho=0.3$ and $\rho=0.7$, under the scenario where the reference sample size was 5,000.  The coverage probabilities are all close to the nominal value of 0.95.  Given the population mean of $y$ is about 9.3, the average lengths of the confidence intervals are also not excessively large.  For example, when $\rho = 0.3$ and $N_B=750$, the 95\% confidence interval for $\hat{\mu}_y$ is $\hat{\mu}_y \pm 1.85$ and the Relative Standard Error (RSE) of the estimate is about 10\%.  When $\rho=0.7$ for the same value of $N_B$, the RSE drops to about 4.6\%.

\begin{table}[ht]
    \centering
    \caption{Coverage Probabilities and Average Length of 95\% Confidence Intervals - Bootstrap variance using EM algorithm}
      \begin{tabular}{lrr}
        \hline
      \multicolumn{1}{c}{} & \multicolumn{2}{c}{$\rho(Y,X\beta)=0.3$} \\
      \multicolumn{1}{c}{$N_B$} & Coverage Probability & Average Length \\
        \hline
        750     &   0.947   &   3.59    \\
        2,000   &   0.941   &   2.09	 \\
        50,000  &   0.969	&   0.32	 \\
        90,000  &   0.970	&   0.23	 \\
        \hline
      \end{tabular}
    \label{tab:coverage_rate_boot}
  \end{table}

Table \ref{tab:coverage_rate_boot} provides the CPs and ALs for the 95\% confidence intervals when we use the bootstrap approach to estimate variances for the OE-EM approach.  Due to the computational time involved, we only produced results for $\rho=0.3$, where the reference sample size was 5,000.  The CPs for the bootstrap approach are again close to the nominal value of 0.95, although it appears that there is a slight over-coverage when $N_B$ is larger.  The ALs again appear to be reasonable relative to the population mean of 9.3, and reduce as sample size increases.

\section{Real-Data Simulation} \label{real_data}

In this section we apply the proposed OE and OE-EM methods alongside the SL, CLW and ALP approaches in a simulation using real-world data where the reference sample may be drawn using sampling approaches other than PPS sampling.

The data we use is based on Australian Business Activity Statement (BAS) data from the July-September quarter of the 2022-2023 financial year.  The BAS data was sourced from the Business Longitudinal Analysis Data Environment (BLADE) standard module and accessed using the Australian Bureau of Statistics (ABS) DataLab.  Records designated as live businesses, in the private or public sector, and with an active Goods and Services Tax (GST) payer role, were extracted.  Pay-As-You-Go (PAYG) data relating to the number of employees in each business was also obtained for the 2022-2023 financial year.  The complete records with no missing data item values - 721,769 records in total - were designated to be our population dataset.

The data item of interest for this exercise was total sales excluding GST (turnover), while available auxiliary items included total salary, wages and other payments (wages), geographic location of the business (state), and a derived industry division classification.  Population distribution across state, as well as population quantiles for turnover and wages, are available in Appendix \ref{app:popn_char}.  In particular, we note that the turnover and wages data items are very right skewed.  The total turnover for the full population during the July-September quarter was \$611,102.1 million, which equated to a mean turnover per business of about \$846,673.

We assessed the various estimation methods by conducting a simulation study using the data from this population.  In each iteration of the simulation, a non-probability dataset was selected using Poisson sampling.  The selection mechanism into the non-probability dataset was driven by a logistic model with one continuous covariate (a standardised wage variable) and one categorical variable (industry division):

\begin{equation}
    \text{log}\Bigl(\frac{\pi^B_i}{1 - \pi^B_i} \Bigr) = \theta_0 + 0.05x_{\text{stdwages}} + \boldsymbol{\theta}^T_{2}\boldsymbol{x}_{\text{industry}} \nonumber
\end{equation}

where $\boldsymbol{\theta}^T_{2} = (0,0,0,0,0,-0.4,-0.1,0,0,0,0,-0.2,0,0,0,0,0,-0.2)$ and $\boldsymbol{x}_{\text{industry}}$ is the industry division variable transformed into 18 0-1 indicator variables (one for each division except for Division A).  These values were chosen to reflect some industries being less likely to be included in the non-probability dataset.  The value $\theta_0$ was chosen so that the size of the non-probability dataset was equal to $100,000$.

We examined the behaviour of the different approaches when the probability sample was chosen using several different sampling schemes.  These included
\begin{enumerate}
    \item \textit{PPS sampling} (PPS): Wages was used as the size variable, capped at \$500,000.  The inclusion probabilities for $\pi^A_i$ were proportional to $z_i = c + \text{wages}_i$, where $c$ was chosen to ensure the ratio max($z_i$)/min($z_i$) was equal to 20.  The probability sample was selected using randomised systematic PPS sampling.
    \item \textit{PPS sampling with probabilities dependent on membership in $B$} (PPS Cond): The same setup as for sampling scheme 1, except that i) the initial selection probabilities for units already in the non-probability dataset were first multiplied by 1.3, and then ii) all probabilities were adjusted so that the sum of the probabilities equalled 1.  These probabilities were  then multiplied by 100,000 to obtain the final selection probability.
    \item \textit{Simple Random Sample} (SRSWOR: A Simple Random Sample Without Replacement (SRSWOR) was drawn from the full population.
    \item \textit{Stratified SRSWOR} (Strat. SRSWOR): The population was assigned to strata based on state, industry division and a size category based on number of employees in a business.  The size categories were: 0 to 4 employees, 5 to 19 employees, 20 to 299 employees, and 300+ employees. An optimal allocation to strata was undertaken using the Bethel-Chromy algorithm (\citealt{bethel_1989}; \citealt{chromy_1987}) to ensure accuracy constraints at the national level (3\%) and industry division level (5\%) were met.  A minimum sample size of 6 in each stratum was applied, while all businesses in the 300+ size strata were completely enumerated.
\end{enumerate}

For the first three sample design scenarios, the sample size for the probability sample was 15,000.  For the Stratified SRSWOR scenario, the total sample size was 14,929.  For each sample design, 2,000 simulations were undertaken for each sample design.  In each iteration of the simulation, the OE, OE-EM, SL, ALP and CLW methods were applied to the drawn samples, and the mean turnover was calculated.

Table \ref{tab:blade_results} provides the Monte Carlo RB and RRMSE values for each estimator under the various sample design scenarios for $A$.  The OE and SL approaches are generally the best performing approaches in all sample design scenarios.  Under the PPS and Stratified SRSWOR design scenarios, the OE and SL RRMSEs are about equal.  However, under the SRSWOR and PPS with conditional probabilities scenarios, the OE approach provides a lower RRMSE.  When the probabilities of selection into $A$ depend on whether or not the unit is in $B$, the SL estimate becomes slightly biased, while the OE approach remains unbiased.

\begin{table}[ht]
    \centering
    \caption{Real-data performance of estimators for turnover}
    \resizebox{0.995\textwidth}{!}{
      \begin{tabular}{crrrrrrrr}
        \hline
        \multicolumn{1}{c}{ } & \multicolumn{2}{c}{PPS} & \multicolumn{2}{c}{PPS Cond} & \multicolumn{2}{c}{SRSWOR} & \multicolumn{2}{c}{Strat. SRSWOR} \\
        \multicolumn{1}{c}{Method} & \%RB & \%RRMSE & \%RB & \%RRMSE & \%RB & \%RRMSE & \%RB & \%RRMSE \\
        \hline
        OE    & -0.39 & 8.47  & -0.57 & 8.53 & -1.35 & 8.95 & 0.25  & 8.60 \\
        OE-EM & 0.46  & 11.36 & 0.57  & 11.15 & -2.14 & 12.85 & 0.21  & 8.56\\
        SL    & -0.24 & 8.41  & 2.06  & 9.38  & -0.33 & 9.19 & 0.23  & 8.58 \\
        CLW   & -1.30 & 13.87  & -0.9  & 13.34 & -7.4  & 16.23 & 0.23  & 8.60\\
        ALP   & 3.06  & 11.36  & 3.22  & 11.38 & 1.41  & 12.32 & 3.96  & 10.97\\
      \hline
      \end{tabular}
    }
        \begin{minipage}{16cm}
			\vspace{0.1cm}
			\small Note: ``OE'' - Overlap-Excluded IPW estimator; ``OE-EM'' - Use of EM algorithm to estimate links between $A$ and $B$; ``SL'' - Sample likelihood approach; ``CLW'' - IPW2 estimator in \citet{chen.etal_2020}; ``ALP'' - ALP estimator in \citet{wang.etal_2021}.
		\end{minipage}
    \label{tab:blade_results}
  \end{table}

The OE-EM approach tends to yield higher RRMSEs compared to the OE and SL approaches, but still tends to outperform the CLW and ALP methods in all four sample design scenarios.  Compared to the other estimators, the ALP approach tends to yield biased results in the simulation under all sample design scenarios.  This bias appears to be due to the fact that the ALP method models a probability $\hat{p}_i = \hat{\pi}^B_i/(1 + \hat{\pi}^B_i)$ (see equation (\ref{eqn:WVL.loglikelihood})), and it turns out that for this population and simulation scenarios the estimates of $\hat{p}_i$ tend to be greater than 0.5.  This means that many of the resulting $\hat{\pi}^B_i = \hat{p}_i/(1 - \hat{p}_i)$ values, where $\hat{p}_i$ is modelled using logistic regression, end up being greater than one.

There is some residual bias in the estimators under the SRSWOR scenario, but this bias is small relative to the RRMSE.  The exception is the CLW approach, which yielded a RB of $-7.4\%$.  The underlying cause for the residual bias appears to be the highly skewed nature of the standardised wages item.  When the (relatively rare) units with the larger wage values are not selected in the simple random sample, the selection bias model coefficient for standardised wages tends to be underestimated, leading to a downward bias in the resulting population estimates.  The OE and SL approaches may be less affected by this due to the fact that they only use the data from the reference sample which does not overlap with the non-probability sample.  To a lesser extent, the OE-EM approach also aims to do this by estimating and then removing the contribution from the overlapping sample.  This result demonstrates the importance of ensuring the reference sample is of sufficient quality so as to effectively assist in correcting selection bias.

\section{Concluding Remarks} \label{conclusion}

In this paper, we have proposed to use knowledge about which reference sample units are also in the non-probability sample to help estimate the participation probabilities of the non-probability units.  The proposed OE pseudo-likelihood excludes from the likelihood expression the probability sample units which are also in the non-probability sample.

We compared how the OE pseudo-likelihood approach performed against several alternative methods in terms of their efficiency when making statistical inferences based on the non-probability sample.  This comparison was undertaken both theoretically, and empirically within a simulation setting using both synthetic data and real-world data.  In the first simulation exercise the OE approach was found to provide lower RRMSE results compared with other pseudo-likelihood based methods when the non-probability datasets were a larger fraction of the population.  Our theoretical comparisons corroborated this empirical finding.  In particular, under the assumption of Poisson sampling for the reference sample, the OE approach was found to always have a lower asymptotic variance compared to the KW and CLW methods, and achieved lower asymptotic variance than the ALP approach when the non-probability sample was larger.

The SL approach was the most efficient approach in the first simulation exercise.  However, we found that in the real-data simulation the efficiency of the SL approach tended to be comparable or slightly worse than the efficiency of the OE approach.  This seems consistent with our theoretical variance comparison, which indicated that SL may not perform as well as OE when the available auxiliary variables only provide a moderate to poor fit and the probabilities $\pi^A_i/(\pi^B_i + \pi^A_i - \pi^A_i\pi^B_i)$ vary only moderately.  When the probability of inclusion into the reference sample depends on its presence in the non-probability dataset, the SL estimate tended to have a slight bias and larger RRMSE, while the OE estimate remained unbiased.  These results suggest that in some practical situations the OE approach may be preferred as the safer option compared to the SL approach which relies on tighter assumptions.

When it is not possible to identify which units in the reference sample are also on the non-probability dataset, the EM-like algorithm proposed in this paper can be used in conjunction with the OE pseudo-likelihood to estimate the non-probability membership status of reference sample units and the model parameters for the propensity scores simultaneously.  Estimating the $\delta_i$ links tends to require a cost in terms of a slightly higher RRMSE, however the empirical studies showed that it can still be preferred compared to other alternatives.  In this paper we did not compare the performance of the EM-like algorithm with the BLUP approach proposed in \citet{beaumont.etal_2024b} when overlap is unknown.  This is something that could be explored as future work.

In this paper we have assumed a SAR missingness mechanism exists in $B$, so that participation probabilities only depend on the auxiliary data items.  We have also assumed that the auxiliary information are captured without measurement error.  In reality, these assumptions may not hold.  Tests can be utilised (see for example \citealt{andridge.etal_2019}) to assess whether a Selection Not At Random (SNAR) situation exists in the non-probability data.  The reference sample units that are in common with the non-probability sample can be used to examine whether there is measurement error in the auxiliary information.  These issues will be explored further in a future paper.


\bigskip

\section*{Supplementary Material}

The derivations relating the common variance expressions for the five methods compared in this article can be found at \textbf{insert link}.

\bigskip

\section*{Acknowledgements}

The first author is grateful to Professor Pengfei Li for sharing R code used for the simulation described in Chen et al (2020).  The first author was supported by funding from the Sir Roland Wilson Foundation and the Australian Bureau of Statistics.

\bigskip

\section*{Appendices}

\renewcommand{\labelenumii}{\textbf{C\arabic{enumii}}}

\renewcommand\thesection{\Alph{section}}
\renewcommand\thesubsection{\thesection.\arabic{subsection}}

\addtocounter{section}{-6}

\subsection{Regularity Conditions for OE Propensities} \label{sec:regularity_conditions}

It is assumed that the following regularity conditions hold:

\begin{enumerate}
	\item[]
    \begin{enumerate}
        \item The finite population size $N$ and the sample sizes $n_A$ and $n_B$ satisfy $\lim_{N \to \infty} n_A/N = f_A \in (0,1)$ and $\lim_{N \to \infty} n_B/N = f_B \in (0,1)$
        \item There exist $c_1$ and $c_2$ such that $0 < c_1 \leq N\pi^A_i/n_A \leq c_2$ and $0 < c_1 \leq N\pi^B_i/n_B \leq c_2$ for all units $i \in \mathcal{U}$.
        \item Under the finite population and the sampling design for $A$, we have $N^{-1}\sum_{i \in A}d^A_i\boldsymbol{u}_i - N^{-1}\sum_{i \in \mathcal{U}}\boldsymbol{u}_i = O_p(n^{-1/2}_A)$ for $\boldsymbol{u}_i=\boldsymbol{x}_i, y_i$, where $d^A_i = 1/\pi^A_i$ and the order in probability is with respect to the sampling mechanism used to select $A$.
        \item The finite population and the propensity scores satisfy $N^{-1}\sum_{i \in \mathcal{U}} y^2_i = O(1)$, $N^{-1}\sum_{i \in \mathcal{U}} \lVert \boldsymbol{x}_i \rVert^3 = O(1)$, and $N^{-1}\sum_{i \in \mathcal{U}}\pi^B_i(1 - \pi^B_i)\boldsymbol{x}_i\boldsymbol{x}_i^T$ is a positive definite matrix.
    \end{enumerate}
\end{enumerate}

Condition \textbf{C1} means that the sample fractions of the probability and non-probability samples are bounded, while condition \textbf{C2} means that the inclusion probabilities for the probability and non-probability samples do not differ in terms of order of magnitude from simple random sampling.  Condition \textbf{C3} ensures consistency of the Horvitz-Thompson estimators obtained from the probability sample.  Condition \textbf{C4} are finite moment conditions to validate Taylor series expansions.

\subsection{Consistency and Variance Derivations for OE Propensities} \label{sec:variance_derivation}

Following \citet{chen.etal_2020}, we set up a joint estimation problem.  Firstly, define $\boldsymbol{\eta}^T = (\mu,\boldsymbol{\theta}^T)$.  The true (target) parameter values are $\boldsymbol{\eta}^T_0 = (\mu_y,\boldsymbol{\theta}^T_0)$, while the OE IPW estimate and propensity score model parameters are combined as $\hat{\boldsymbol{\eta}}^T = (\hat{\mu},\hat{\boldsymbol{\theta}}^T)$.  $\hat{\boldsymbol{\eta}}^T$ is the solution to the combined estimating equation system

\begin{align}
    \boldsymbol{\Phi}(\boldsymbol{\eta}) &= 
        \begin{pmatrix}
            U(\mu,\boldsymbol{\theta}) \\
            S(\boldsymbol{\theta})
        \end{pmatrix} \nonumber \\
        &=
        \begin{pmatrix}
            \frac{1}{N}\sum_{i \in \mathcal{U}}\frac{\delta_i(y_i-\mu)}{\pi(\boldsymbol{x};\boldsymbol{\theta})} \\
            \frac{1}{N}\sum_{i \in \mathcal{U}} \Bigl\{\delta_i\boldsymbol{x}_i - \delta_i\pi(\boldsymbol{x};\boldsymbol{\theta})\boldsymbol{x}_i \Bigr\} - \frac{1}{N}\sum_{i \in \mathcal{U}}I_i(1 -  \delta_i)d^A_i\pi(\boldsymbol{x};\boldsymbol{\theta})\boldsymbol{x}_i
        \end{pmatrix}
        = \boldsymbol{0}. \nonumber
\end{align}

We consider the joint randomisation of the propensity score model and the sample design for $A$.  Recall that $\pi^B_i = \pi(\boldsymbol{x}_i; \boldsymbol{\theta}_0)$.  When Assumptions \textbf{A1}, \textbf{A2}, and the propensity score model all hold, and further assuming that the probability sample is drawn independent of the non-probability sample, then $E\{\boldsymbol{\Phi}(\boldsymbol{\eta}_0)\} = \boldsymbol{0}$, since
\begin{align}
    \frac{1}{N}\sum_{i \in \mathcal{U}}\frac{E(\delta_i)(y_i-\mu_y)}{\pi^B_i} = \frac{1}{N}\sum_{i \in \mathcal{U}}(y_i-\mu_y) = 0 \nonumber
\end{align}

and
\begin{align}
    &\frac{1}{N}\sum_{i \in \mathcal{U}} \Bigl\{E(\delta_i)\boldsymbol{x}_i - E(\delta_i)\pi^B_i\boldsymbol{x}_i \Bigr\} - \frac{1}{N}\sum_{i \in \mathcal{U}}(E(I_i) -  E(I_i\delta_i))d^A_i\pi^B_i\boldsymbol{x}_i \nonumber \\
    &= \frac{1}{N}\sum_{i \in \mathcal{U}} \Bigl\{ \pi^B_i \boldsymbol{x}_i - (\pi^B_i)^2 \boldsymbol{x}_i \Bigr\} - \frac{1}{N}\sum_{i \in \mathcal{U}}(\pi^A_i -  \pi^A_i\pi^B_i)\frac{1}{\pi^A_i}\pi^B_i\boldsymbol{x}_i \nonumber \\
    &= \boldsymbol{0} \nonumber
\end{align}

Note that $\boldsymbol{\Phi}(\boldsymbol{\hat{\eta}}) = \boldsymbol{0}$.  Under the regularity conditions in Appendix \ref{sec:regularity_conditions}, we apply a first-order Taylor expansion to obtain
\begin{align} \label{eqn:tse}
    \boldsymbol{0} = \boldsymbol{\Phi}(\boldsymbol{\hat{\eta}}) \approx \boldsymbol{\Phi}(\boldsymbol{\eta}_0) + \boldsymbol{\phi}(\boldsymbol{\eta}_0)(\hat{\boldsymbol{\eta}} - \boldsymbol{\eta}_0)
\end{align}

where $\boldsymbol{\phi}(\boldsymbol{\eta})$ is defined as $\partial \boldsymbol{\Phi}(\boldsymbol{\eta})/\partial \boldsymbol{\eta}$, and is given by

\begin{equation}
    \boldsymbol{\phi}(\boldsymbol{\eta}) =
        \begin{pmatrix}
            \frac{\partial U}{\partial \mu} & \frac{\partial U}{\partial \boldsymbol{\theta}} \\
            \frac{\partial S}{\partial \mu} & \frac{\partial S}{\partial \boldsymbol{\theta}}
        \end{pmatrix} \nonumber
\end{equation}

Under suitable regularity conditions as $n \to \infty$ (see \citealt{stefanski.boos_2002}), we have
\begin{align}
    \frac{1}{n}\sum_{i=1}^{n}\frac{\partial \boldsymbol{\Phi}(\boldsymbol{x}_i,y_i,\boldsymbol{\eta})}{\partial \boldsymbol{\eta}} \overset{p}{\to} E \Bigl\{ \frac{\partial \boldsymbol{\Phi}(\boldsymbol{x},y,\boldsymbol{\eta})}{\partial \boldsymbol{\eta}} \Bigr\} \nonumber
\end{align}

We can then re-arrange (\ref{eqn:tse}) to obtain
\begin{align}
    \hat{\boldsymbol{\eta}} - \boldsymbol{\eta}_0 &\approx -[\boldsymbol{\phi}(\boldsymbol{\eta}_0)]^{-1}\boldsymbol{\Phi}(\boldsymbol{\eta}_0) \nonumber \\
    &\approx -[E(\boldsymbol{\phi}(\boldsymbol{\eta}_0))]^{-1}\boldsymbol{\Phi}(\boldsymbol{\eta}_0) \label{eqn_app:asympt_eta}
\end{align}

Now $E\{\boldsymbol{\Phi}(\boldsymbol{\eta}_0)\} = \boldsymbol{0}$, and drawing on the standard M-estimation argument of \citet{stefanski.boos_2002}, we can conclude that $E(\hat{\boldsymbol{\eta}})$ converges to $\boldsymbol{\eta}_0$ asymptotically, that is, $\hat{\boldsymbol{\eta}} \overset{p}{\to} {\boldsymbol{\eta}}_0$ as the population and sample sizes tend to infinity.

To derive the variance for the OE IPW estimator, we apply the sandwich formula (\citealt{stefanski.boos_2002}):
\begin{equation}
    Var(\hat{\boldsymbol{\eta}}) = E\{\boldsymbol{\phi}(\boldsymbol{\eta}_0)\}^{-1}Var\{\boldsymbol{\Phi}(\boldsymbol{\eta}_0)\}[E\{\boldsymbol{\phi}(\boldsymbol{\eta}_0)\}^{-1}]^{T} \label{eqn:sandwich}
\end{equation}

The components of $\boldsymbol{\phi}(\boldsymbol{\eta})$ can be derived giving
\begin{equation}
    \boldsymbol{\phi}(\boldsymbol{\eta}) =
        \begin{pmatrix}
            \frac{-1}{N}\sum_{i \in \mathcal{U}}\frac{\delta_i}{\pi(\boldsymbol{x}_i; \boldsymbol{\theta})} & \frac{-1}{N}\sum_{i \in \mathcal{U}}\frac{1 - \pi(\boldsymbol{x}_i; \boldsymbol{\theta})}{\pi(\boldsymbol{x}_i; \boldsymbol{\theta})}\delta_i(y_i-\mu)\boldsymbol{x}^T_i \\
            \boldsymbol{0} & \frac{-1}{N}\sum_{i \in \mathcal{U}}\bigl\{ \delta_i + I_i(1-\delta_i)d^A_i\bigr\} \pi(\boldsymbol{x}_i; \boldsymbol{\theta})(1 - \pi(\boldsymbol{x}_i; \boldsymbol{\theta}))\boldsymbol{x}_i\boldsymbol{x}^T_i
        \end{pmatrix} \nonumber
\end{equation}

Then
\begin{align}
    E\{\boldsymbol{\phi}(\boldsymbol{\eta}_0)\} &=
        \begin{pmatrix}
            -1 & \frac{-1}{N}\sum_{i \in \mathcal{U}}(1 - \pi^B_i)(y_i-\mu_y)\boldsymbol{x}^T_i \\
            \boldsymbol{0} & \frac{-1}{N}\sum_{i \in \mathcal{U}}\pi^B_i(1 - \pi^B_i)\boldsymbol{x}_i\boldsymbol{x}^T_i
        \end{pmatrix} \nonumber
\end{align}

and the inverse is given as
\begin{align}
    E\{\boldsymbol{\phi}(\boldsymbol{\eta}_0)\}^{-1}
    &=
        \begin{pmatrix}
            -1 & \sum_{i \in \mathcal{U}}(1 - \pi^B_i)(y_i - \mu_y)\boldsymbol{x}^T_i \Bigl[ \sum_{i \in \mathcal{U}}\pi^B_i(1 - \pi^B_i)\boldsymbol{x}_i\boldsymbol{x}^T_i \Bigr]^{-1} \\
            \boldsymbol{0} & -\Bigl[ \frac{1}{N}\sum_{i \in \mathcal{U}}\pi^B_i(1 - \pi^B_i)\boldsymbol{x}_i\boldsymbol{x}^T_i \Bigr]^{-1} \label{eqn:inv_exp_phi}
        \end{pmatrix}.
\end{align}

To determine $Var\{\boldsymbol{\Phi}_n(\boldsymbol{\eta}_0)\}$, we calculate $Var(U(\mu_y,\boldsymbol{\theta}_0))$, $Var(S(\boldsymbol{\theta}_0))$, and $Cov(U(\mu_y,\boldsymbol{\theta}_0),S(\boldsymbol{\theta}_0))$.

Firstly, $Var(U(\mu_y,\boldsymbol{\theta}_0))$ is derived as
\begin{align}
    Var(U(\mu_y,\boldsymbol{\theta}_0)) &= \frac{1}{N^2}\sum_{i \in \mathcal{U}}\frac{Var(\delta_i)(y_i - \mu_y)^2}{(\pi^B_i)^2} \nonumber \\
    &= \frac{1}{N^2}\sum_{i \in \mathcal{U}}\frac{(1 - \pi^B_i)(y_i - \mu_y)^2}{\pi^B_i} \label{eqn:var_u}
\end{align}

To find $Var(S(\boldsymbol{\theta}_0))$, we make use of the decomposition
\begin{align}
    Var(S(\boldsymbol{\theta}_0)) &= E_d\Bigl\{Var(S(\boldsymbol{\theta}_0)|A) \Bigr\} + Var_d\Bigl\{ E(S(\boldsymbol{\theta}_0)|A) \Bigr\} \nonumber
\end{align}
where $E_d$ and $Var_d$ refer to expectations and variances with respect to the probability sample design respectively.

Now
\begin{align}
    Var(S(\boldsymbol{\theta}_0)|A) &= Var \Bigl[ \frac{1}{N}\sum_{i \in \mathcal{U}} \Bigl\{\delta_i\boldsymbol{x}_i - \delta_i\pi^B_i\boldsymbol{x}_i \Bigr\} - \frac{1}{N}\sum_{i \in \mathcal{U}}I_i(1 -  \delta_i)d^A_i\pi^B_i\boldsymbol{x}_i | A \Bigr] \nonumber \\
    &= Var \Bigl[ \frac{1}{N}\sum_{i \in \mathcal{U}} \delta_i(1 - \pi^B_i + d^A_iI_i\pi^B_i) \boldsymbol{x}_i - \frac{1}{N}\sum_{i \in \mathcal{U}}I_id^A_i\pi^B_i\boldsymbol{x}_i | A \Bigr] \nonumber \\
    &= \frac{1}{N^2}\sum_{i \in \mathcal{U}} (1 - \pi^B_i + d^A_iI_i\pi^B_i)^2\boldsymbol{x}_i\boldsymbol{x}^T_i Var(\delta_i) \nonumber
\end{align}

Expanding out and taking the expectation, we obtain
\begin{align}
    E_d\Bigl\{Var(S(\boldsymbol{\theta}_0)|A) \Bigr\} 
    &= \frac{1}{N^2}\sum_{i \in \mathcal{U}}\Bigl[ \pi^B_i(1 - \pi^B_i)(1 + (d^A_i - 1)(\pi^B_i)^2) \boldsymbol{x}_i\boldsymbol{x}^T_i \Bigr]. \label{eqn:exp_var}
\end{align}

Then
\begin{align}
    E(S(\boldsymbol{\theta}_0)|A) &= \frac{1}{N}\sum_{i \in \mathcal{U}}\Bigl[ \pi^B_i\boldsymbol{x}_i - (\pi^B_i)^2\boldsymbol{x}_i \Bigr] + \frac{1}{N}\sum_{i \in A}(\pi^B_i)^2d^A_i\boldsymbol{x}_i - \frac{1}{N}\sum_{i \in A}d^A_i\pi^B_i\boldsymbol{x}_i \nonumber \\
    \implies Var_d\Bigl\{ E(S(\boldsymbol{\theta}_0)|A) \Bigr\} &= \frac{1}{N^2}Var_d\Bigl[ \sum_{i \in A}\{(\pi^B_i)^2d^A_i\boldsymbol{x}_i - d^A_i\pi^B_i\boldsymbol{x}_i \} \Bigr] \nonumber \\
    &= \frac{1}{N^2}Var_d\Bigl(\sum_{i \in A}d^A_i\pi^B_i(1 - \pi^B_i)\boldsymbol{x}_i \Bigr) \label{eqn:var_exp}
\end{align}

where $Var_d(\boldsymbol{x})$ is the design-based variance-covariance matrix for variables $\boldsymbol{x}$ under the probability sample design.

Combining (\ref{eqn:exp_var}) and (\ref{eqn:var_exp}), we obtain
\begin{align}
    Var(S(\boldsymbol{\theta}_0)) = \frac{1}{N^2}\sum_{i \in \mathcal{U}}\Bigl[ \pi^B_i(1 - \pi^B_i)(1 + (d^A_i - 1)(\pi^B_i)^2) \boldsymbol{x}_i\boldsymbol{x}^T_i \Bigr] + \frac{1}{N^2}Var_d\Bigl(\sum_{i \in A}d^A_i\pi^B_i(1 - \pi^B_i)\boldsymbol{x}_i \Bigr) \label{eqn:var_s}
\end{align}

To find $Cov(U(\mu_y,\boldsymbol{\theta}_0),S(\boldsymbol{\theta}_0))$, we will use
\begin{align}
    Cov(U(\mu_y,\boldsymbol{\theta}_0),S(\boldsymbol{\theta}_0)) = E_d\Bigl[Cov(U(\mu_y,\boldsymbol{\theta}_0),S(\boldsymbol{\theta}_0)|A) \Bigr] + Cov_d\Bigl[E(U(\mu_y,\boldsymbol{\theta}_0)|A),E(S(\boldsymbol{\theta}_0)|A) \Bigr] \nonumber
\end{align}
where $E_d$ is defined as previously and $Cov_d$ is the covariance under the probability sample design.

Now,
\begin{align}
    E(U(\mu_y,\boldsymbol{\theta}_0)|A) &= \frac{1}{N}\sum_{i \in \mathcal{U}}(y_i - \mu_y) = 0 \nonumber \\
    &\implies Cov_d\Bigl[E(U(\mu_y,\boldsymbol{\theta}_0)),E(S(\boldsymbol{\theta}_0)|A) \Bigr] = 0. \label{eqn:cov_exp}
\end{align}

Finally,
\begin{align}
    &Cov(U(\mu_y,\boldsymbol{\theta}_0),S(\boldsymbol{\theta}_0)|A) \nonumber \\
    &= Cov\Bigl[ \frac{1}{N}\sum_{i \in \mathcal{U}}\frac{\delta_i(y_i - \mu_y)}{\pi^B_i}, \frac{1}{N}\sum_{i \in \mathcal{U}} \{\delta_i\boldsymbol{x}_i - \delta_i\pi^B_i\boldsymbol{x}_i - I_i(1 - \delta_i)d^A_i\pi^B_i\boldsymbol{x}_i\} |A \Bigr] \nonumber \\
    &= \frac{1}{N^2}\sum_{i \in \mathcal{U}}Cov\Bigl[\frac{\delta_i(y_i - \mu_y)}{\pi^B_i},\delta_i\{\boldsymbol{x}_i - \pi^B_i\boldsymbol{x}_i + I_id^A_i\pi^B_i\boldsymbol{x}_i\}|A \Bigr] \nonumber \\
    &= \frac{1}{N^2}\sum_{i \in \mathcal{U}}(y_i - \mu_y)(1 - \pi^B_i)(1 - \pi^B_i + I_id^A_i\pi^B_i)\boldsymbol{x}_i \nonumber \\
    &\implies E_d\Bigl[Cov(U(\mu_y,\boldsymbol{\theta}_0),S(\boldsymbol{\theta}_0)|A) \Bigr] = \frac{1}{N^2}\sum_{i \in \mathcal{U}}(y_i - \mu_y)(1 - \pi^B_i)\boldsymbol{x}_i \label{eqn:exp_cov}
\end{align}
where we make use of the bilinear property of the covariance function and assume that $\delta_i \perp \delta_j$ for $i,j \in B$.

Combining (\ref{eqn:cov_exp}) and (\ref{eqn:exp_cov}), we get
\begin{align}
    Cov(U(\mu_y,\boldsymbol{\theta}_0),S(\boldsymbol{\theta}_0)) = \frac{1}{N^2}\sum_{i \in \mathcal{U}}(y_i - \mu_y)(1 - \pi^B_i)\boldsymbol{x}_i \label{eqn:cov}
\end{align}

Denote the components of $E\{\boldsymbol{\phi}(\boldsymbol{\eta}_0)\}^{-1}$ and $Var\{\boldsymbol{\Phi}(\boldsymbol{\eta}_0)\}$ as
\begin{equation}
    E\{\boldsymbol{\phi}(\boldsymbol{\eta}_0)\}^{-1} = 
    \begin{pmatrix}
        -1 & \boldsymbol{b}^T \\
        \boldsymbol{0} & \boldsymbol{d}
    \end{pmatrix},
    Var\{\boldsymbol{\Phi}(\boldsymbol{\eta}_0)\} = 
    \begin{pmatrix}
        e & \boldsymbol{f} \\
        \boldsymbol{f} & \boldsymbol{h}
    \end{pmatrix} \nonumber
\end{equation}

where
\begin{align}
    \boldsymbol{b}^T &= \sum_{i \in \mathcal{U}}(1 - \pi^B_i)(y_i - \mu_y)\boldsymbol{x}^T_i \Bigl[ \sum_{i \in \mathcal{U}}\pi^B_i(1 - \pi^B_i)\boldsymbol{x}_i\boldsymbol{x}^T_i \Bigr]^{-1} \text{ from (}\ref{eqn:inv_exp_phi}\text{)} \nonumber \\
    \boldsymbol{d} &= -\Bigl[ \frac{1}{N}\sum_{i \in \mathcal{U}}\pi^B_i(1 - \pi^B_i)\boldsymbol{x}_i\boldsymbol{x}^T_i \Bigr]^{-1} \nonumber \text{ from (}\ref{eqn:inv_exp_phi}\text{)} \\
    e &= \frac{1}{N^2}\sum_{i \in \mathcal{U}}\frac{(1 - \pi^B_i)(y_i - \mu_y)^2}{\pi^B_i} \text{ from (}\ref{eqn:var_u}\text{)} \nonumber \\
    \boldsymbol{f} &= \frac{1}{N^2}\sum_{i \in \mathcal{U}}(y_i - \mu_y)(1 - \pi^B_i)\boldsymbol{x}_i \text{ from (}\ref{eqn:cov}\text{)} \nonumber \\
    \boldsymbol{h} &= \frac{1}{N^2}\sum_{i \in \mathcal{U}}\Bigl[ \pi^B_i(1 - \pi^B_i)(1 + (d^A_i - 1)(\pi^B_i)^2) \boldsymbol{x}_i\boldsymbol{x}^T_i \Bigr] + \frac{1}{N^2}Var_d\Bigl(\sum_{i \in A}d^A_i\pi^B_i(1 - \pi^B_i)\boldsymbol{x}_i \Bigr) \nonumber \\
    &\text{ from (}\ref{eqn:var_s}\text{)} \nonumber
\end{align}

Then, using the sandwich formula, $Var(\hat{\boldsymbol{\eta}})$ can be derived as
\begin{align} \label{eqn:varIPW}
    Var(\hat{\boldsymbol{\eta}}) &= E\{\boldsymbol{\phi}(\boldsymbol{\eta}_0)\}^{-1}Var\{\boldsymbol{\Phi}(\boldsymbol{\eta}_0)\}[E\{\boldsymbol{\phi}(\boldsymbol{\eta}_0)\}^{-1}]^{T} \nonumber \\
    &=
    \begin{pmatrix}
        -1 & \boldsymbol{b}^T \\
        \boldsymbol{0} & \boldsymbol{d}
    \end{pmatrix}
    \begin{pmatrix}
        e & \boldsymbol{f}^T \\
        \boldsymbol{f} & \boldsymbol{h}
    \end{pmatrix}
    \begin{pmatrix}
        -1 & \boldsymbol{0}^T \\
        \boldsymbol{b} & \boldsymbol{d}
    \end{pmatrix} \nonumber \\
    &=
    \begin{pmatrix}
        e - \boldsymbol{b}^T\boldsymbol{f}-\boldsymbol{b}^T\boldsymbol{f}+\boldsymbol{b}^T\boldsymbol{h}\boldsymbol{b} & & - \boldsymbol{f}^T\boldsymbol{d}+\boldsymbol{b}^T\boldsymbol{h}\boldsymbol{d} \\
        -\boldsymbol{d}\boldsymbol{f}+\boldsymbol{d}\boldsymbol{h}\boldsymbol{b} & & \boldsymbol{d}\boldsymbol{h}\boldsymbol{d} 
    \end{pmatrix} 
\end{align}

In particular, the variance of the OE IPW estimator is the first diagonal element of the matrix (\ref{eqn:varIPW}):
\begin{align}
    Var(\hat{\mu}) &=  e - \boldsymbol{b}^T\boldsymbol{f}-\boldsymbol{b}^T\boldsymbol{f}+\boldsymbol{b}^T\boldsymbol{h}\boldsymbol{b} \nonumber \\
    &= \frac{1}{N^2}\sum_{i \in \mathcal{U}} \frac{1-\pi^B_i}{\pi^B_i}(y_i-\mu_y)^2 - \frac{\boldsymbol{b}^T}{N^2}(y_i-\mu_y)(1 - \pi^B_i)\boldsymbol{x}_i - \nonumber \\
    &\mathrel{\phantom{=}} \frac{\boldsymbol{b}^T}{N^2}\sum_{i \in \mathcal{U}}(y_i - \mu_y)(1 - \pi^B_i)\boldsymbol{x}_i + \boldsymbol{b}^T\boldsymbol{h}\boldsymbol{b} \nonumber \\
    &= \frac{1}{N^2}\sum_{i \in \mathcal{U}} (1-\pi^B_i)\pi^B_i \Bigl[\frac{(y_i - \mu_y)^2}{(\pi^B_i)^2} - 2\frac{(y_i - \mu_y)\boldsymbol{b}^T\boldsymbol{x}_i}{\pi^B_i}   \Bigr] + \boldsymbol{b}^T\boldsymbol{h}\boldsymbol{b} \nonumber \\
    &= \frac{1}{N^2}\sum_{i \in \mathcal{U}} (1-\pi^B_i)\pi^B_i \Bigl[\frac{(y_i - \mu_y)^2}{(\pi^B_i)^2} - 2\frac{(y_i - \mu_y)\boldsymbol{b}^T\boldsymbol{x}_i}{\pi^B_i}   \Bigr] + \nonumber \\
       &\mathrel{\phantom{=}} \boldsymbol{b}^T \frac{1}{N^2}\sum_{i \in \mathcal{U}}\Bigl[\pi^B_i(1 - \pi^B_i)\Bigl\{1 + (d^A_i - 1)(\pi^B_i)^2 \Bigr\} \boldsymbol{x}_i\boldsymbol{x}^T_i \Bigr] \boldsymbol{b} + \boldsymbol{b}^T\boldsymbol{V}\boldsymbol{b} \nonumber \\
    &= \frac{1}{N^2}\sum_{i \in \mathcal{U}} (1-\pi^B_i)\pi^B_i \Bigl[\frac{(y_i - \mu_y)^2}{(\pi^B_i)^2} - 2\frac{(y_i - \mu_y)}{\pi^B_i}\boldsymbol{b}^T\boldsymbol{x}_i + \boldsymbol{b}^T\boldsymbol{x}_i\boldsymbol{x}^T_i\boldsymbol{b}  \Bigr] + \nonumber \\
    &\mathrel{\phantom{=}}  \boldsymbol{b}^T \frac{1}{N^2}\sum_{i \in \mathcal{U}}\Bigl[\pi^B_i(1 - \pi^B_i)\Bigl\{(d^A_i - 1)(\pi^B_i)^2 \Bigr\} \boldsymbol{x}_i\boldsymbol{x}^T_i \Bigr] \boldsymbol{b} + \boldsymbol{b}^T\boldsymbol{V}\boldsymbol{b}\nonumber \\
    &= \frac{1}{N^2}\sum_{i \in \mathcal{U}} (1-\pi^B_i)\pi^B_i \Bigl[\frac{(y_i - \mu_y)}{(\pi^B_i)} - \boldsymbol{b}^T\boldsymbol{x}_i  \Bigr]^2 + \label{eqn:varIPW CLW term tech} \\
    &\mathrel{\phantom{=}} \boldsymbol{b}^T\boldsymbol{V}\boldsymbol{b} \text{ } +  \label{eqna:varIPW desvar} \\
    &\mathrel{\phantom{=}} \frac{1}{N^2}\sum_{i \in \mathcal{U}}\Bigl[(\pi^B_i)^3(1 - \pi^B_i)(d^A_i - 1) \boldsymbol{b}^T\boldsymbol{x}_i\boldsymbol{x}^T_i\boldsymbol{b} \Bigr] \label{eqna:varIPW addterm}
\end{align}

where $\boldsymbol{V} = \frac{1}{N^2} Var_d \Bigl(\sum_{i \in A}d^A_i\pi^B_i(1-\pi^B_i)\boldsymbol{x}_i \Bigr)$.

\subsection{Additional simulation results} \label{app:additional_results}

\begin{table}[H]
    \centering
    \caption{Comparative performance of estimators - $n_A=5,000$ and $\rho(y,\boldsymbol{x}^T\beta)=0.7$}
      \begin{tabular}{lrrrrrrrr}
        \hline
      \multicolumn{1}{c}{} & \multicolumn{2}{c}{$N_B=750$} & \multicolumn{2}{c}{$N_B=2,000$}
        & \multicolumn{2}{c}{$N_B=50,000$} & \multicolumn{2}{c}{$N_B=140,000$} \\
       & \%RB & \%RRMSE & \%RB & \%RRMSE & \%RB & \%RRMSE & \%RB & \%RRMSE \\
        \hline
        Naive & 85.06 & 85.09 & 73.70 & 73.71 & 27.62 & 27.62 & 8.13  & 8.13 \\
        OE    & -0.20 & 4.44  & -0.16 & 2.65  & -0.02 & 0.52  & -0.01 & 0.27 \\
        OE-EM & -0.06 & 4.35  & -0.06 & 2.64  & 0.00  & 0.58  & 0.00  & 0.38 \\
        SL    & 0.04  & 3.97  & 0.02  & 2.26  & 0.00  & 0.40  & -0.01 & 0.26 \\
        CLW   & -0.41 & 5.13  & -0.26 & 3.33  & 0.24  & 0.97  & -0.06 & 0.72 \\
        ALP   & 0.84  & 3.97  & 1.01  & 2.43  & 0.45  & 0.69  & 0.04  & 0.50 \\
        KW    & 0.15  & 7.85  & -0.10 & 4.46  & -0.05 & 0.97  & -0.03 & 0.38 \\
      \hline
      \end{tabular}
      \begin{minipage}{16cm}
			\vspace{0.1cm}
			\small Note: ``Naive'' - Simple sample mean from $B$; ``OE'' - Overlap-Excluded IPW estimator; ``OE-EM'' - Use of EM algorithm to estimate links between $A$ and $B$; ``SL'' - Sample likelihood approach; ``CLW'' - IPW2 estimator in \citet{chen.etal_2020}; ``ALP'' - ALP estimator in \citet{wang.etal_2021}; ``KW'' - Estimator proposed by \citet{kim.wang_2019}.
	   \end{minipage}
      \label{tab:sim2_results}
\end{table}

\begin{table}[H]
    \centering
    \caption{Comparative performance of estimators - $n_A=500$ and $\rho(y,\boldsymbol{x}^T\beta)=0.7$}
      \begin{tabular}{lrrrrrrrr}
        \hline
      \multicolumn{1}{c}{} & \multicolumn{2}{c}{$N_B=750$} & \multicolumn{2}{c}{$N_B=2,000$}
        & \multicolumn{2}{c}{$N_B=50,000$} & \multicolumn{2}{c}{$N_B=100,000$} \\
       & \%RB & \%RRMSE & \%RB & \%RRMSE & \%RB & \%RRMSE & \%RB & \%RRMSE \\
        \hline
        Naive & 85.06 & 85.09 & 73.70 & 73.71 & 27.62 & 27.62 & 14.97 & 14.97 \\
        OE & -2.24 & 6.97  & -1.23 & 5.02  & -0.13 & 1.54  & -0.05 & 1.08 \\
        OE-EM & -1.05 & 6.99  & -0.49 & 4.88  & -0.05 & 1.79  & 0.03  & 1.40 \\
        SL & 0.06  & 3.96  & 0.05  & 2.48  & 0.02  & 1.14  & 0.01  & 0.94 \\
        CLW & -4.76 & 11.51 & -2.68 & 8.41  & -0.68 & 3.93  & -0.69 & 3.58 \\
        ALP & 0.35  & 3.97  & 0.37  & 2.52  & 0.11  & 1.62  & 0.06  & 1.61 \\
      \hline
      \end{tabular}
      \begin{minipage}{16cm}
			\vspace{0.1cm}
			\small Note: ``Naive'' - Simple sample mean from $B$; ``OE'' - Overlap-Excluded IPW estimator; ``OE-EM'' - Use of EM algorithm to estimate links between $A$ and $B$; ``SL'' - Sample likelihood approach; ``CLW'' - IPW2 estimator in \citet{chen.etal_2020}; ``ALP'' - ALP estimator in \citet{wang.etal_2021}.
	   \end{minipage}
      \label{tab:sim2_results_small}
\end{table}

\subsection{Selected population characteristics for real-world data application} \label{app:popn_char}
\begin{table}[h!]
    \centering
    \caption{Population distribution across state}
    \resizebox{0.995\textwidth}{!}{
      \begin{tabular}{lrrrrrrrr}
        \hline
        & NSW & VIC & QLD & SA & WA & TAS & NT & ACT \\
        \hline
        Business Count & 249,089 & 187,065   & 142,994   & 43,229    & 68,727    & 13,830    & 5,529 & 11,306 \\
        \hline
      \end{tabular}
    }
      \begin{minipage}{15.5cm}
		\vspace{0.1cm}
        \spacingset{1}
		\small Note: Derived population counts created using BLADE indicative items and BAS data from Quarter 1 2022-2023, as well as PAYG data from 2022-2023 (\cite{businesslongitudinalanalysisdataenvironmentblade_2022}) and BAS data from Quarter 1 2022-2023, accessed through the ABS DataLab December 2024.
	   \end{minipage}
    \label{tab:blade_state}
  \end{table}

\begin{table}
    \spacingset{1}
    \centering
    \caption{Selected quantiles for turnover and wages}
      \begin{tabular}{lrrrrr}
        \hline
         & \multicolumn{5}{c}{Quantile} \\
        & 1\% & 25\% & 50\% & 75\% & 99\% \\
        \hline
        Turnover & \$0 & \$59,408 & \$147,527 & \$403,741 & \$10,101,149 \\
        Wages & \$0 & \$15,150 & \$39,108 & \$102,078 & \$2,074,381 \\
        \hline
      \end{tabular}
      \begin{minipage}{10.5cm}
		\vspace{0.1cm}
        \spacingset{1}
		\small Note: Derived quantiles created using BLADE indicative items and BAS data from Quarter 1 2022-2023, as well as PAYG data from 2022-2023 (\cite{businesslongitudinalanalysisdataenvironmentblade_2022}) and BAS data from Quarter 1 2022-2023, accessed through the ABS DataLab December 2024.
	   \end{minipage}
    \label{tab:blade_quantiles}
  \end{table}

\newpage

\section*{ABS DataLab Disclaimer}

The data used to create the derived population for the business turnover application in this paper was supplied to the ABS under the Taxation Administration Act 1953, A New Tax System (Australian Business Number) Act 1999, Australian Border Force Act 2015, Social Security (Administration) Act 1999, A New Tax System (Family Assistance) (Administration) Act 1999, Paid Parental Leave Act 2010 and/or the Student Assistance Act 1973. Such data may only be used for the purpose of administering the Census and Statistics Act 1905 or performance of functions of the ABS as set out in section 6 of the Australian Bureau of Statistics Act 1975. No individual information collected under the Census and Statistics Act 1905 is provided back to custodians for administrative or regulatory purposes. Any discussion of data limitations or weaknesses is in the context of using the data for statistical purposes and is not related to the ability of the data to support the Australian Taxation Office, Australian Business Register, Department of Social Services and/or Department of Home Affairs’ core operational requirements.

Legislative requirements to ensure privacy and secrecy of these data have been followed. For access to PLIDA and/or BLADE data under Section 16A of the ABS Act 1975 or enabled by section 15 of the Census and Statistics (Information Release and Access) Determination 2018, source data are de-identified and so data about specific individuals has not been viewed in conducting this analysis. In accordance with the Census and Statistics Act 1905, results have been treated where necessary to ensure that they are not likely to enable identification of a particular person or organisation.

\printbibliography

\end{document}